\documentstyle{mn}
\input psfig

\def\ltsim{\lower.5ex\hbox{$\; \buildrel < \over \sim \;$}}
\def\gtsim{\lower.5ex\hbox{$\; \buildrel > \over \sim \;$}}

\title[Biasing and the distribution of dark matter haloes]
{Biasing and the distribution of dark matter haloes}

\author[Sheth \& Lemson]
{Ravi K. Sheth$^1$ \& Gerard Lemson$^{1,2,3}$ \\
$^1$ Max-Planck Institut f\"ur Astrophysik, Karl-Schwarzschild-Str. 1,
85740 Garching, Germany\\
$^2$ Racah Institute of Physics, The Hebrew University, Jerusalem, 
91904 Israel\\
$^3$ 271 Dartmouth Street, Apt. 4F, Boston, MA 02116, USA\\
\smallskip
Email: sheth@mpa-garching.mpg.de, jkimlemson@erols.com\\}

\date{Accepted 1998 December; Submitted 1998 November 10; 
in original form 1998 August 13}
 
\begin{document}

\maketitle

\begin{abstract}
In hierarchical models of gravitational clustering, virialized 
haloes are biased tracers of the matter distribution.  
As discussed by Mo \& White (1996), this bias is nonlinear and 
stochastic.  They developed a model which allows one to write down 
analytic expressions for the mean of the bias relation, in the 
initial Lagrangian, and the evolved, Eulerian, spaces.  We provide 
analytic expressions for the higher order moments as well.  

In the initial Lagrangian space, each halo occupies a volume that 
is proportional to its mass.  Haloes cannot overlap initially, so 
this gives rise to volume exclusion effects which can have 
important consequences for the halo distribution, particularly on 
scales smaller than that of a typical halo.  
Our model allows one to include these volume exclusion 
effects explicitly when computing the mean and higher order 
statistics of the Lagrangian space halo distribution.  
As a result of dynamical evolution, the spatial distribution of 
haloes in the evolved Eulerian space is likely to be different from 
that in the initial Lagrangian space.  When combined with the 
Mo \& White spherical collapse model, the model developed here allows 
one to quantify the evolution of the mean and scatter of the bias 
relation.  We also show how their approach can be extended to 
compute the evolution, not just of the haloes, but of the dark 
matter distribution itself.  

Biasing and its evolution depend on the initial power spectrum.  
Clustering from Poisson and white noise Gaussian initial conditions 
is treated in detail, since, in these cases, exact analytical 
results are available.  We conjecture that these results can
be easily extended to provide an approximate but accurate model 
for the biasing associated with clustering from more general 
Gaussian initial conditions.  For all initial power spectra 
studied here, the model predictions for the Eulerian 
bias relation are in reasonable agreement with numerical simulations 
of hierarchical gravitational clustering for haloes of a wide range 
of masses, whereas the predictions for the corresponding Lagrangian 
space quantities are accurate only for massive haloes.
\end{abstract}


\section{Introduction}
In hierarchical models of gravitational clustering, it is possible 
to use the statistical properties of the initial density field, 
assumed to be Gaussian, to compute good approximations to the 
average number density of virialized objects at subsequent times 
(Press \& Schechter 1974).  
In this paper, the number density of virialized objects will be 
called the unconstrained mass function.  
The statistical properties of the initial dark matter distribution 
can also be used to compute merger models which describe some aspects 
of how virialized haloes at a late time were assembled by mergers of 
smaller ones which, themselves, had virialized earlier (Bond et al. 
1991).  For example, the average number of $M_1$ haloes identified 
at $t_1$ that merged to form an $M_0$ halo by time $t_0$ can be 
computed (Lacey \& Cole 1993, 1994).  In this paper, this quantity 
will be called the constrained mass function.  
Associated with any given object is a merger history tree which 
describes how the object was assembled.  An analytic model that 
describes the merger trees of dark matter haloes has only been 
developed for the special case of Poisson initial conditions 
(Sheth 1996).  With some care, it can also be used to describe 
the merger trees of haloes identified in white noise initial 
conditions (Sheth \& Pitman 1997; Sheth \& Lemson 1998).  

In all these analyses, the number density of haloes was computed, 
but their spatial distribution was not.  Recently, Mo \& White (1996) 
described a model which uses the initial dark matter distribution to 
estimate the initial Lagrangian space distribution of dark matter 
haloes.  Dynamical evolution is likely to modify this distribution, 
so that the distribution in the final Eulerian space is  
different from that initially.  Mo \& White also formulated a model 
for this evolution.  In their model, statistical quantities in the 
Eulerian space are obtained by transforming the corresponding 
Lagrangian space quantities appropriately.  In their model, then, 
the problem is to compute the Lagrangian space quantities, since, 
once these are known, the corresponding Eulerian quantities 
follow trivially.  

In the Mo \& White model, haloes are biased tracers of the 
underlying matter distribution, the bias between haloes and 
mass being, in general, nonlinear and stochastic.  They showed 
that, on average, the bias relation depends only on the constrained 
and unconstrained mass functions, but that knowledge of the 
higher order moments of the merger history tree is required 
to compute the scatter around this mean correctly.  Since they did 
not have an analytic model for the merger history tree, they were 
able to obtain analytic results for the scatter in the bias relation, 
or for the halo--halo correlation function, only in the limit of 
large separations.  In this limit, the mean bias relation is linear, 
and the scatter around this relation is Poisson.   

Since a halo in the Lagrangian space occupies a volume that is 
proportional to its mass, and since haloes do not overlap, the 
Lagrangian space halo distribution is a particular case of a 
hard-sphere model.  As Mo \& White discuss, the associated 
volume exclusion effects will introduce anti-correlations on scales 
smaller than that of a typical halo.  On these scales, the scatter 
in the bias relation may well be less than Poisson.  
This paper combines some of the ideas contained in Mo \& White (1996) 
with the analytic merger model of Sheth (1996) to provide a 
description of the evolution of the higher order moments of the halo 
distribution that incorporates these exclusion effects explicitly.  
Thus, within the context of the Mo--White model, the results 
presented here are valid even on the small scales where the mean 
bias relation is nonlinear.  

Although the analytic merger tree described by Sheth (1996) was 
derived for the special case of Poisson initial conditions, 
it also describes the trees associated with white noise Gaussian 
initial conditions (e.g. Sheth \& Pitman 1997).  
Sheth \& Lemson (1998) showed that it could be used to derive 
reasonably accurate analytic approximations to the higher order 
moments of the merger tree distribution associated with more general 
Gaussian initial conditions.  When combined with the Mo \& White 
model, this allows us to write down analytic approximations for the 
higher order moments of, e.g. the bias relation, for more general 
Gaussian initial conditions, that should also be reasonably accurate.  

This paper is organized as follows.  
The Lagrangian space halo distribution associated with white noise 
initial conditions is described in Section~\ref{wnics}.  
This section also serves to set notation.  
The white noise results are extended to describe the Lagrangian 
space halo distribution in more general Gaussian random fields 
in Section~\ref{scfics}.  
Section~\ref{scmow} contains a brief summary of the Mo \& White 
spherical collapse model for computing Eulerian space quantities 
given the corresponding Lagrangian ones.  It also shows how the 
model can be extended to compute the Eulerian space probability 
distribution function of the matter as well as the haloes.  
Section~\ref{sims} shows the results of comparing the model 
predictions with the distribution of haloes identified in numerical 
simulations of gravitational clustering.  This section also 
compares the model predictions for the stochasticity of the bias 
relation with what is measured in the simulations.  
A final section summarizes our results.  

All the Lagrangian space results of this paper follow from results 
originally derived for haloes which form from Poisson initial 
conditions.  Since these initial conditions are unfamiliar to most 
readers, the description of clustering from Poisson initial 
conditions is given in an Appendix.  
The Poisson case has the virtue that everything can be worked 
out rigorously, so readers interested in the various subtle 
issues involved in this approach are encouraged to read it.  

\section{White-noise initial conditions}\label{wnics}
This section provides a description of the initial halo 
distribution when the initial matter distribution is a 
white-noise Gaussian random field.  
Sections~\ref{mfs}--\ref{meanbias} summarize various 
known results.  They are included to set notation, and to 
clarify the logic that leads to the final expressions.  
Section~\ref{homg} provides analytic expressions for 
the higher order moments of the Lagrangian space halo 
distribution.  These moments are related to the higher 
order moments of the bias relation, and are the principal 
new results of this paper.  

\subsection{Unconditional and conditional mass functions}\label{mfs}
To set notation it is useful to summarize various known results.  
Assume that the initial density field $\delta$ is Gaussian, 
with power spectrum $P(k)$.  If the field is smoothed with a 
spherically symmetric filter of size $V$, then the smoothed field 
$\delta(V)$ is also Gaussian.  This means that the one point 
probability distribution function is 
\begin{equation}
p(\delta,V)\,{\rm d}\delta= {1\over \sqrt{2\pi S}}\,
\exp\left(-{\delta^2\over 2S}\right)\ {\rm d}\delta ,
\label{g0}
\end{equation}
where $S\equiv \langle \delta(V)^2 \rangle$.  That is, 
\begin{equation}
S \equiv {1\over (2\pi)^2}\ 
\int_0^\infty 4\pi k^2\,P(k)\,W^2(kR)\ {\rm d}k ,
\label{svar}
\end{equation}
where $W$ is the Fourier transform of the smoothing 
window, and $V\propto R^3$ with the constant of proportionality 
depending on the shape of the window.  In this section we will 
mainly be concerned with a window which is a top hat in real 
space, for which $W(x) = (3/x^3)[\sin(x) - x\,\cos(x)]$, 
and $V = 4\pi R^3/3$.  

Let $\bar\rho$ denote the average background density.  
If $P(k)\propto k^n$, then $S\propto (\bar\rho V)^{-\alpha}$, 
where $\alpha = (n+3)/3$.  
If $n=0$ the random field is said to be white noise.  
The mass contained within the filter is 
$M\equiv \bar\rho V(1+\delta)$.  
Notice that when $S\ll 1$, then $|\delta|\ll 1$ almost surely.  
In this case, $\delta<-1$ is extremely unlikely, 
so there is no problem with defining the mass as was done above.  

We will assume that $S\ll 1$ in the initial conditions, which we 
will sometimes call the Lagrangian space.  Then, in Lagrangian 
space, $|\delta|\ll 1$, so to lowest order in $\delta$, 
$M\equiv \bar\rho V$, and $S\propto M^{-\alpha}$.  
We will always be concerned with initial Gaussian fluctuation 
fields for which the relation between $S$ and $V$, and so 
the relation between $S$ and $M$ is monotonic.  
Thus, in Lagrangian space, $M$, $S$ and $V$ are all equivalent 
variables.  

Most of the expressions associated with the excursion set 
approach concern properties of Gaussian random fields when they 
are smoothed on different scales.  Here we will assume that 
the filter is a top hat in real space, and that the initial 
Lagrangian space distribution is Gaussian white noise.  
For white noise $\alpha=1$, so $S = (\bar\rho V)^{-1}$, and 
the conditional probability that the field has value $\delta_1$ 
when smoothed on scale $V_1$, given that it had value $\delta_0$ 
when smoothed on scale $V_0$ is  
\begin{equation}
p(\delta_1,V_1|\delta_0,V_0) = 
{1\over\sqrt{2\pi(S_1-S_0)}}\ 
\exp\left(-{(\delta_1-\delta_0)^2\over 2(S_1-S_0)}\right) .
\label{g10}
\end{equation}
Let $q(\delta_1,\delta_0,V_0)$ denote the probability that, 
when smoothed on scale $V_0$, the density is $\delta_0$, and 
that it is less dense than $\delta_1$ for all $V>V_0$.  Then 
\begin{equation}
q(\delta_1,\delta_0,V_0) = p(\delta_0,V_0)
\left[1-\exp\left(-{2\delta_1(\delta_1-\delta_0)\over S_0}\right)\right],
\label{qchandra}
\end{equation}
provided $\delta_1>\delta_0$, and it is equal to $0$ otherwise 
(e.g. Chandrasekhar 1943).  Of course, this means that 
$q<p$ as expected.  

In the excursion set approach, virialized dark matter haloes are 
associated with isolated regions:  these are those Lagrangian regions 
that, when smoothed on some scale $V$ are denser than some critical 
density, and when smoothed on still larger scales, are less dense 
than this (Bond et al. 1991).  
All the mass contained within this critically overdense 
isolated $V$ is associated with a virialized halo.  
This required critical density is a function of time, but not of 
smoothing scale $V$.  It decreases with increasing time:  
haloes that virialize at late times are associated with less dense 
isolated regions in Lagrangian space than haloes which virialize at 
early times.  Let $\delta_{\rm c}(z)$ denote this critical density, 
and let $f(M,\delta_{\rm c})\,{\rm d}M$ denote the fraction of 
Lagrangian space that is taken up by volumes $V$ that have density 
$\delta_{\rm c}(z)$ when smoothed on scale $V$, and are less dense 
on all larger scales, so that each such isolated $V$ is associated 
with a halo of mass $M$ that has just virialized at the epoch 
labelled by $z$.  In Lagrangian space $S$, the mass $M$, 
and the associated volume $V$ are all equivalent variables, so 
$f(M,\delta_{\rm c})\,{\rm d}M=f(S,\delta_{\rm c})\,{\rm d}S$, 
and 
\begin{equation}
f(S,\delta_{\rm c})\,{\rm d}S = 
{1\over \sqrt{2\pi}}\,{\delta_{\rm c}\over S^{3/2}}\,
\exp\left(-{\delta_{\rm c}^2\over 2S}\right)\,{\rm d}S 
\label{fm}
\end{equation}
(Bond et al. 1991).  
The associated number density of such isolated regions is the 
same as the number density of virialized objects, and is given by 
\begin{equation}
n(M,\delta_{\rm c})\,{\rm d}M = 
{\bar\rho\over M}\ {f(S,\delta_{\rm c})\,{\rm d}S} = 
\bar\rho\ {f(M,\delta_{\rm c})\,{\rm d}M \over M}.
\label{nm1}
\end{equation}
This is sometimes called the unconstrained, or universal 
mass function (Press \& Schechter 1974).  
Now, since $S$ and $M$ are equivalent variables, the integral 
of $f(S,\delta)$ over all $S$ is the same as the integral of 
$f(M,\delta)$ over all $M$.  Equation~(\ref{fm}) shows that 
this integral is unity.  This can be interpretted as showing that 
associated with any given epoch $z$ is a partition of the total 
Lagrangian volume into isolated regions of volume $V$ and overdensity 
$\delta_{\rm c}(z)$; the mass in each region $V$ first virializes 
to form a halo of mass $M=\bar\rho V$ at $z$.  

Now consider some $\delta_1\ge \delta_0$, where $\delta_1$ is 
a convenient notation for $\delta_{\rm c}(z_1)$, and we have 
assumed that $z_1>z_0$, so $z$ increases with decreasing epoch.  
Restrict attention to Lagrangian regions $V_0$ that 
are associated with $M_0$ haloes at the epoch $z_0$;  i.e., isolated 
regions $V_0$.  Consider one such isolated region.  
Suppose that when smoothed on the scale $V_1\le V_0$ this region 
is denser than $\delta_1$, and that it is less dense than this for 
all larger smoothing scales.  Then $V_1$ is an isolated subregion 
within $V_0$; this isolated Lagrangian subregion $V_1$ within $V_0$ 
can be associated with a subhalo $M_1$ of $M_0$; $M_1$ will first 
virialize at the epoch $z_1$.  Let $f(M_1|M_0)\,{\rm d}M_1$ denote 
the fraction of the mass of $M_0$ that, at the epoch $z_1$, is 
associated with subclumps $M_1$.  Since $S$ and $M$ are equivalent 
variables, $f(M_1|M_0)\,{\rm d}M_1 = f(S_1|S_0)\,{\rm d}S_1$ where 
\begin{eqnarray}
f(S_1,\delta_1|S_0,\delta_0)\,{\rm d}S_1 &=& 
{1\over \sqrt{2\pi}}\,{(\delta_1-\delta_0)\over(S_1-S_0)^{3/2}}\nonumber \\
& &\ \ \times\exp\left[-{(\delta_1-\delta_0)^2\over 2(S_1-S_0)}\right]
\,{\rm d}S_1 
\label{fm1m2}
\end{eqnarray}
(Bond et al. 1991; Lacey \& Cole 1993).  
Integrating this over the range $0\le M_1\le M_0$ gives 
unity:  all the mass of $M_0$ was in subclumps of some smaller 
mass at the earlier epoch $z_1>z_0$.  
This fraction can be converted into a mean number of $M_1$ haloes 
within an $M_0$ halo:  
\begin{equation}
{\cal N}(M_1,\delta_1|M_0,\delta_0) = 
{M_0\over M_1}\,f(M_1,\delta_1|M_0,\delta_0) .
\label{n10}
\end{equation}
Since $M_0=\bar\rho V_0$, we should divide ${\cal N}(1|0)$ by $V_0$ 
to express it as a number density.  Then comparison with 
equation~(\ref{nm1}) shows why this expression is sometimes called 
the constrained mass function.  
Equation~(\ref{n10}) can also be understood as follows.  
For any given $z_1>z_0$, the mass $M_0$ contained within an 
isolated Lagrangian region $V_0$ within which the average density 
is $\delta_0$, so the region first virializes at $z_0$, 
can be thought of as being partitioned into isolated subregions, 
each of slightly higher density $\delta_1$.  

\subsection{The first moment of the Lagrangian space halo distribution}
The previous expressions mean that the mean number of 
$(M_1,\delta_1)$ haloes that are in randomly placed Lagrangian 
cells of size $V_0$ is $n(M_1,\delta_1)\,V_0$.  
Let $\bar N(M_1,\delta_1|\delta_0,V_0)$ denote the average number 
of $(M_1,\delta_1)$ haloes in a Lagrangian cell $V_0$ that has 
overdensity $\delta_0$.  Then, by definition, 
\begin{equation}
n(M_1,\delta_1)\,V_0 = \int_{-\infty}^\infty 
\bar N(M_1,\delta_1|\delta_0,V_0)\ p(\delta_0,V_0)\ {\rm d}\delta_0 .
\label{nbarh}
\end{equation}
Since mass and volume are equivalent variables, we will assume 
that $\bar N(1|0)=0$ if $M_1>M_0$.  
Below, we show that when $M_1\le M_0$, then $\bar N(1|0)$ is 
related to ${\cal N}(1|0)$, and that equation~(\ref{nbarh}) is 
consistent with the results of the previous subsection.  

Classify all cells $V_0$ by the overdensity within them.  
Each cell with density $\delta_0$ is either isolated or not.  
By definition, cells with $\delta_0>\delta_1$ are not isolated.  
For cells that are not isolated, $\bar N(1|0)=0$.
Since there is no contribution from cells that are not isolated, 
to compute the average number of $(M_1,\delta_1)$ haloes, we now 
need to sum up the contribution from cells that are isolated.  

If isolated, a cell can be partitioned into isolated subregions 
that are identified with $\delta_1$ haloes.  Label each such 
partition ${\bmath m}$, where ${\bmath m}$ lists the mass associated 
with each subregion.  Let $\pi({\bmath m})$ denote the set of all 
such partitions, and let $p({\bmath m}|M_0)$ denote the probability 
of having the particular partition ${\bmath m}$ (we have not written 
explicilty that this probability will also depend on $\delta_1$ and 
$\delta_0$).  We must integrate over all partitions 
${\bmath m}$ of $(M_0,\delta_0)$ haloes, sum up the number 
$n(M_1|{\bmath m},M_0)$ of $(M_1,\delta_1)$ haloes in each 
partition, weight by the probability $p({\bmath m}|M_0)$ that that 
partition occured, and then integrate over all values of $\delta_0$, 
weighting by the probability that $V_0$ with density $\delta_0$ is 
isolated.  The sum over partitions gives the average number of 
equation~(\ref{n10}):  
\begin{equation}
\bar N(1|0) = \int_{\pi({\bmath m})} 
n({M_1|\bmath m},M_0)\ p({\bmath m}|M_0) = {\cal N}(1|0) ,
\end{equation}
where the final equality follows because the integral is over 
all partitions ${\bmath m}$ of $M$, so it is the definition of 
${\cal N}(1|0)$.  This means that 
\begin{equation}
n(M_1,\delta_1)\,V_0 = \int_{-\infty}^{\delta_1} 
{\cal N}(1|0)\ q(\delta_1,\delta_0,V_0)\ {\rm d}\delta_0 ,
\label{feq10}
\end{equation}
where the fact that only isolated cells give a nonzero contribution 
to the integral in equation~(\ref{nbarh}) means that the upper limit 
in the integral over $\delta_0$ must be $\delta_1$, and that we must 
replace $p(\delta_0,V_0)$ with $q(\delta_1,\delta_0,V_0)$, the 
fraction of cells of density $\delta_0$ that are isolated.  
Simple algebra shows that equations~(\ref{n10}) and~(\ref{qchandra}), 
when substituted into the right hand side of this expression, 
do satisfy this relation.  

The main reason for writing this out explicitly is that it 
shows how one might begin to quantify the extent to which virialized 
haloes are biased tracers of the underlying matter distribution.  
We do this in the next section.  

\subsection{The mean bias relation and the cross correlation 
between haloes and mass}\label{meanbias}
Let 
\begin{equation}
\Delta_{\bmath m}(1|0) = 
{n(M_1|{\bmath m},M_0)\over n(M_1,\delta_1)V_0} - 1 
\end{equation}
denote the average overdensity of $M_1$ haloes within an $M_0$ 
halo that is known to be partitioned into the haloes ${\bmath m}$.  
Integrating this over all partitions gives 
\begin{eqnarray}
\delta_{\rm h}^{\rm L}(1|0) &\equiv& 
\int_{\pi({\bmath m})} 
{\rm d}\Delta_{\bmath m}\ \Delta_{\bmath m}(1|0)\,p({\bmath m}|M_0)
\nonumber \\
&=& {{\cal N}(M_1,\delta_1|M_0,\delta_0)\over n(M_1\delta_1)V_0} - 1.
\label{dhl10}
\end{eqnarray}
This gives the mean overdensity of $(M_1,\delta_1)$ haloes that 
are within $(M_0,\delta_0)$ haloes.  It can also be understood as 
the mean overdensity of isolated $(M_1,\delta_1)$ regions that are 
within isolated $(M_0,\delta_0)$ regions in the Lagrangian space.  
In regions that are not isolated, (e.g., if $\delta_0>\delta_1$) 
$\delta_{\rm h}^{\rm L}=-1$.  
Thus, $\delta_{\rm h}^{\rm L}(1|0)$ is the same as the mean 
bias relation of equation~(12) in Mo \& White (1996).  
The peak background split (their equation~13) is obtained 
in the limit in which the cell size $V_0$ is much larger 
than the Lagrangian size of an $M_1$ halo 
(e.g. Bardeen et al. 1986), 
\begin{equation}
\delta_{\rm h}^{\rm L}(1|0) \to 
{\nu_1^2 - 1\over \delta_1}\ \delta_0\equiv B(1|0)\,\delta_0 ,
\label{pbsplit}
\end{equation}
where $\nu_1^2 = \delta_1^2/S_1$, and the final equality 
defines $B(1|0)$.  

Notice that the mean overdensity of the halo distribution is a 
linear function of the mass overdensity only in the limit of 
equation~(\ref{pbsplit}).  Equation~(\ref{dhl10}) 
shows that, in general, this mean bias relation is nonlinear.  
Just as the mean bias relation depends on the mean number of 
haloes in Lagrangian cells $(M_0,\delta_0)$, 
the higher order moments of the Lagrangian bias relation depend 
on the higher order moments of the Lagrangian space halo 
distribution.  We will compute these higher order moments 
in the next subsection.  If these higher order moments are 
nonzero, then there will be some scatter around this mean bias 
relation:  in addition to being nonlinear, the bias will be 
stochastic.  

Before doing so, we will first calculate the Lagrangian space 
cross correlation between haloes and mass, averaged over all 
randomly placed Lagrangian cells $V_0$.  This is 
\begin{eqnarray}
\bar\xi_{\rm hm}^{\rm L}(M_1,\delta_1|V_0) &\equiv&
\Bigl\langle 
\left({\bar N(1|0)\over n(M_1,\delta_1)V_0}-1\right)\,\delta_0
\Bigr\rangle
\nonumber \\
&=& \int_{-\infty}^\infty \!\!
{\bar N(1|0)\over n(M_1,\delta_1)V_0}\,\delta_0\ 
p(\delta_0,V_0)\,{\rm d}\delta_0.
\label{xihmdef}
\end{eqnarray}
In the first line, the integral is over all Lagrangian cells, so 
the second equality follows since $\langle\delta_0 \rangle\equiv 0$.  
This integral is the sum of two terms, the first due to those 
Lagrangian cells that are isolated, and the second due to those that 
are not.  However, $\bar N(1|0)=0$ for cells that are not isolated.  
For isolated cells, the contribution is computed by 
a double average, one over all values of $\delta_0$ with 
the substitution $p(\delta_0)\to q(\delta_0)$, 
and the other over all partitions of ${\bmath m}$.  The 
integral over partitions gives $\bar N(1|0) = {\cal N}(1|0)$, so 
\begin{equation}
\bar\xi_{\rm hm}^{\rm L}(M_1,\delta_1|V_0) = 
\int_{-\infty}^{\delta_1} 
{{\cal N}(1|0)\over n(M_1,\delta_1)V_0}\,\delta_0\ 
q(\delta_1,\delta_0,V_0)\ {\rm d}\delta_0 .
\label{xihml10}
\end{equation}
This expression for the cross correlation between haloes and mass 
is the same as equation~(15) in Mo \& White (1996), but 
with a difference in interpretation.  As we have shown, the average 
is to be understood as being over all randomly placed Lagrangian 
cells $V_0$, not just those that are less dense than $\delta_1$.  

The integral in~(\ref{xihml10}) can be done analytically:
\begin{eqnarray}
{\bar\xi_{\rm hm}^{\rm L}(1|0)\over S_0} \!\!&=&\!\!
{\delta_1\over S_0} - 
{(\nu^2_{10}+1)\over \delta_1}\ {\rm erf}\left(\nu_{10}\over\sqrt{2}\right)
 -\sqrt{{2\nu_{10}^2\over\pi}}\ {{\rm e}^{-\nu^2_{10}/2}\over\delta_1},
\nonumber \\
\!\!\!\!{\rm where}\!\!&&\!\!
\nu^2_{10} = {\delta_1^2\,(S_1-S_0)\over S_0S_1} .
\end{eqnarray}
When $S_0\ll 1$, then the error function tends to unity and 
the third term tends to zero.  Thus, 
\begin{equation}
{\bar\xi_{\rm hm}^{\rm L}(1|0)\over S_0} \to 
{1\over \delta_1}
\left({\delta^2_1\over S_0} - \nu^2_{10} - 1\right) = B(1|0).
\end{equation}
This is consistent with using equation~(\ref{pbsplit}) for 
$\delta_{\rm h}^{\rm L}(1|0)$ in equation~(\ref{dhl10}) 
and substituting in~(\ref{xihmdef}).  

\subsection{Higher order moments of the bias relation and 
halo halo correlations}\label{homg}
Suppose that there are $n$ $M_1$ haloes within an $M_0$ halo.  
The Lagrangian volume associated with these haloes is $n\,V_1$.  
The average overdensity of the remaining volume is   
\begin{equation}
1+\delta^{(n)} = {M_0 - nM_1\over V_0 - nV_1},\qquad
{\rm where}\ \delta^{(0)} = \delta_0 .
\end{equation}
Since $M_1=V_1(1+\delta_1)$, 
\begin{equation}
\delta_1-\delta^{(n)} = (\delta_1-\delta_0)\ {M_0\over M_0 - nM_1}
\label{deltans}
\end{equation}
to lowest order in the $\delta$ terms.  With this definition, 
the $i$th factorial moment is 
\begin{equation}
\phi_i(M_1,\delta_1|M_0,\delta_0)\ = 
\prod_{n=0}^{i-1} {\cal N}\Bigl(M_1,\delta_1|M_0-nM_1,\delta^{(n)}\Bigr),
\label{mualpha}
\end{equation}
provided $iM_1\le M_0$, and it is zero otherwise.  
This formula is essentially a reworking of results originally in 
Sheth (1996).  See Appendix~\ref{pics} of this paper or 
Sheth \& Lemson (1998) for details.  Following the same 
logic as for the mean (the case $i=1$), the $i$th factorial moment 
of the corresponding halo counts in cells distribution is 
\begin{eqnarray}
&&\!\!\!\!\!\!\!\!\!\!\!\!\!
\int_{-\infty}^{\delta_1} \phi_i(M_1,\delta_1|M_0,\delta_0)\ 
q(\delta_1,\delta_0,V_0)\ {\rm d}\delta_0  \nonumber \\
&&\equiv\ \Bigl(n(M_1,\delta_1)\,V_0\Bigr)^i 
\Bigl(1 + \Xi_i(M_1,\delta_1,V_0)\Bigr) ,
\label{muaqd0}
\end{eqnarray}
where the final equality defines $\Xi_i(M_1,\delta_1,V_0)$.
If the scatter of halo counts were Poisson, then $\Xi_i=0$.  
For $i>1$, equation~(\ref{muaqd0}) can be solved analytically.  
For example, when $i$ is even, then it reduces to a sum of 
incomplete Gamma functions.  Thus, it is possible to show 
explicitly that the scatter is not Poisson.  

Recall that the scatter in the bias relation is related to 
the higher order moments of the halo distribution.  
For example, the variance in the bias relation is essentially 
the same as the variance in the halo distribution.  
In general, this variance is neither zero, nor is it the 
same as the mean.  In other words, the mean bias is nonlinear, 
it is stochastic, and the rms scatter around the mean is not the 
canonical square-root-of-the-mean value that is typical of a 
Poisson distribution.  
To see why, we turn now to a more detailed study of the halo halo 
correlation functions.  

Define 
\begin{eqnarray}
\omega\equiv {\nu_{\rm rem}\over 2s_0},\ \ {\rm where}\ 
\nu_{\rm rem} \equiv 1 - {S(M_0)\over S(iM_1)}\ {\rm and}\ 
{1\over s_0} \equiv {\delta_1^2\over S_0}, 
\label{vrem}
\end{eqnarray}
with $S_0 \equiv S(M_0)$.  
These two parameters have simple physical interpretations.  
An $(M_1,\delta_1)$-halo occupies a volume $V_1$ in the 
initial Lagrangian space, and, by assumption, all the mass 
within $V_1$ is associated with $M_1$.  That is, haloes are 
spatially exclusive; they do not overlap with other.  If 
a randomly placed $V_0$ contains $i$ haloes, each of 
initial size $V_1$, then $\nu_{\rm rem}$ is related to the 
fraction of $V_0$ that is not occupied by these haloes.  
Since $M_0\propto V_0$, $s_0$ expresses the cell size $V_0$ 
in units of the (Lagrangian) size of typical haloes at time 
$\delta_1$, since the usual definition of a typical $M_*$ 
halo is that $\delta^2/S_*\equiv\delta^2/S(M_*)\equiv 1$.  

For white noise, $\Xi_i(M_1,\delta_1,V_0)$ is not a function of 
$M_1$, $\delta_1$, and $V_0$, individually, but only of 
$\nu_{\rm rem}$ and $s_0$.  Thus, $\Xi_i(M_1,\delta_1,V_0)$ has a 
self-similar form; for haloes defined at a given $\delta_1$, it 
depends only on the cell size relative to the size of typical 
objects with the same $\delta_1$, and on the size of the objects 
being measured relative to the cell size.  
Let $\bar\xi_{\rm hh}^{\rm L}(11|0)\equiv \Xi_2(M_1,\delta_1,V_0)$.  
Note that this means that $\bar\xi_{\rm hh}^{\rm L}(11|0)$ denotes 
the volume average of the halo-halo correlation function.  It 
is related to $\xi_{\rm hh}^{\rm L}$ itself by the relation 
\begin{displaymath}
\bar\xi_{\rm hh}^{\rm L}(11|0) = {3\over R_0^3}\,
\int_0^{\rm R_0} \xi_{\rm hh}^{\rm L}(r)\ r^2\,{\rm d}r .
\label{volavg}
\end{displaymath}
This volume average is the variance of halo counts in Lagrangian 
cells of size $V_0$ divided by the square of the mean number of 
halo counts, minus the shotnoise contribution, $1/M_0$ which 
accounts for the fact that the haloes are discrete objects.  
Equation~(\ref{muaqd0}) implies that 
\begin{equation}
1 + \bar\xi_{\rm hh}^{\rm L}(11|0) = {2s_0\over\sqrt{\pi}}
\left[ \sqrt{\omega}\, {\rm e}^{-\omega} +
       [\omega + 0.5]\,\gamma\left(0.5,\omega\right) \right].
\label{xihh}
\end{equation}

If $c(M_1,M_2,\delta_1|M_0,\delta_0)$ denotes the cross correlation 
between $M_1$ and $M_2$ haloes, each with initial overdensity 
$\delta_1$, that are both within the same $M_0$ halo of initial 
overdensity $\delta_0$, then the same logic that led to 
equation~(\ref{mualpha}) implies that 
\begin{eqnarray}
&&\!\!\!\!\!\!\!\!\!\!\!\!\!\!\!\!\!
c(M_1,M_2,\delta_1|M_0,\delta_0) \equiv  \nonumber \\
&& \ {\cal N}(M_1,\delta_1|M_0,\delta_0)\ 
{\cal N}(M_2,\delta_1|M_0-M_1,\delta^{(1)}),
\label{fcross}
\end{eqnarray}
where $\delta^{(1)}$ was defined earlier (equation~\ref{deltans}).  
The volume averaged cross correlation function is got by averaging 
$c(12|0)$ over all isolated volumes $V_0$:  
\begin{eqnarray}
&&\!\!\!\!\!\!\!\!\!\!\!\!
1+\bar\xi_{\rm hh}^{\rm L}(12|0)\ = \nonumber \\
&&\!\!\!\!\!\!\!\!\!\!\!\! 
\int_{-\infty}^{\delta_1} 
{c(M_1,M_2,\delta_1|M_0,\delta_0)\over 
n(M_1,\delta_1)V_0\ n(M_2,\delta_1)V_0 }\, 
q(\delta_1,\delta_0,V_0)\ {\rm d}\delta_0 .
\label{c120q}
\end{eqnarray}
Thus, $\bar\xi_{\rm hh}^{\rm L}(12|0)$ is given by an expression 
that is exactly like equation~(\ref{xihh}), except that now 
$\nu_{\rm rem} = (S_{12} - S_0)/S_{12}$, with 
$S_{12} = S(M_1+M_2)$.  
For white noise, the actual values of $S_1$ and $S_2$ are 
unimportant, only $S_{12}$ matters; given $M<M_0$, 
$\bar\xi_{\rm hh}^{\rm L}(M_1,M-M_1,\delta_1|V_0)$ is the same for 
all values of $M_1<M$.  This suggests that when 
$\bar\xi_{\rm hh}^{\rm L}(12|0)$ differs from zero, 
it is because of volume exclusion effects only.  

\begin{figure}
\centering
\mbox{\psfig{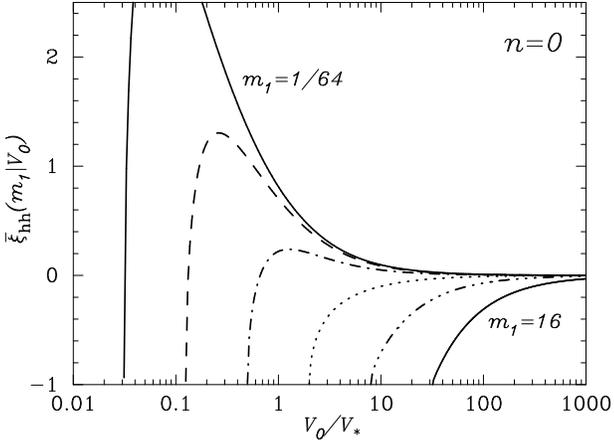}}
\caption{The volume average of the halo-halo correlation function, 
$\bar\xi_{\rm hh}^{\rm L}(11|0)$, given by equation~(\ref{xihh}), 
as a function of cell size $V_0$, for white noise initial conditions.  
The different curves are for haloes of mass 
$m_1=1/64, 1/16, 1/4, 1, 4$ and $16$, respectively.  For white 
noise initial conditions, this is also a plot of the average 
cross-correlation between haloes whose mass sums to $2\,m_1$.}
\label{xihh0}
\end{figure}
Figure~\ref{xihh0} shows $\bar\xi_{\rm hh}^{\rm L}(11|0)$ as a 
function of cell size $V_0$ for white noise initial conditions.  
The different curves show a range of choices of the halo mass $m_1$.  
Masses and scales are in units of the characteristic mass 
$M_*$ and scale $V_*=M_*/\bar\rho$, respectively.  
For white noise initial conditions, this is also a plot of the 
average cross-correlation between haloes whose mass 
sums to $2\,m_1$.  

The shapes of these curves are easily understood.  
Consider haloes that have the same mass $M$.  
Given this mass, there are three scales in the problem:  
the Lagrangian scale of each halo, $V$, 
the initial mean separation between such haloes, $R$, and 
the Lagrangian scale associated with a typical $M_*$ halo, $V_*$.  
Let $m\equiv M/M_*$, $v\equiv V/V_*$, $W\propto R^3$ and $
w\equiv W/V_*$.  
Equation~(\ref{nm1}) shows that the number density of less massive 
($m\ll 1$) haloes is $\propto m^{-3/2}/V_*$.  
The mean separation volume $W$ is the inverse of this, so $w<1$.  
The number density of massive haloes ($m\gg 1$) decreases 
exponentially.  For these haloes $w>1$.

Now, by definition all haloes are anticorrelated on scales smaller 
than that which they occupy (since it takes two haloes to make a 
pair, this scale is $2V$).  
Massive haloes have $w>1$.  Since the mean separation between 
such haloes is large, they are not affected by the fact that 
some of the volume is excluded.  Suppose that, on scales larger 
than $2V$, these haloes were uncorrelated with each other.  
Then $\xi_{\rm hh}^{\rm L}=-1$ on small scales, and 
$\xi_{\rm hh}^{\rm L}=0$ on larger scales, so that on large scales 
the volume average is  $\bar\xi_{\rm hh}^{\rm L}(m|V_0)\propto -2V/V_0$.  
This gives approximately the same qualitative behaviour as the 
limiting relation~(\ref{xilim}).  Namely, 
$\bar\xi_{\rm hh}^{\rm L}$ is always negative, and it becomes less 
negative with increasing scale $V_0$.  

Less massive haloes have $w<1$.  These haloes are affected by 
the excluded volume, since a large fraction of the volume they 
could have occupied is now excluded.  This means that they must 
all be crowded into the remaining volume, so over a range of 
scales, they will appear to be correlated with each other.  
Thus, for white noise initial conditions, volume exclusion 
produces two effects.  
Firstly, all haloes are anti-correlated on scales smaller than 
that which they occupy.  
Secondly, less massive haloes are positively correlated on 
intermediate scales, whereas more massive haloes are 
essentially uncorrelated on all scales larger than those 
which they occupy.  
Thus, on small scales, and for less massive haloes, volume 
exclusion gives rise to effects which are in the opposite 
sense to the commonly held view that less massive haloes are 
also less correlated.  

\subsection{The large-volume limit}
Before moving on to consider more general initial conditions 
than white-noise, it is useful to write down the large scale 
limits of the Lagrangian space halo-halo correlation functions.  

When $M_0\gg (M_1 + M_2)$ and $\delta_0<\delta_1$, then use 
of the asymptotic expansion of the error function reduces 
equation~(\ref{xihh}) to 
\begin{equation}
\bar\xi_{\rm hh}^{\rm L}(12|0)\to 
\left[B(1|0)\,B(2|0)-{\nu_1^2\,\nu_2^2\over\delta_1^2}\right]\,S_0 ,  
\label{xilim}
\end{equation} 
where $\nu_2^2$ and $B(2|0)$ are defined similarly to $\nu_1^2$ and 
$B(1|0)$ (cf. equation~\ref{pbsplit}).  Since the factor 
$(\nu_1\nu_2/\delta_1)^2$ is not necessarily small, this limiting 
form shows that volume exclusion effects are important, even on 
large scales, for massive haloes.  In fact, in this limit 
$\bar\xi_{\rm hh}^{\rm L}(12|0)/S_0\to (1-\nu_1^2-\nu_2^2)/\delta_1^2$, 
so massive haloes are less clustered than less massive haloes on 
all scales.  

For $n>2$, define 
\begin{equation}
H_n \equiv {\bar\xi_n\over\bar\xi_2^{n-1}},
\qquad{\rm where}\ \bar\xi_2 \equiv \bar\xi_{\rm hh}^{\rm L},
\end{equation}
and $\bar\xi_n$ denotes the volume average of the $n$-point 
Lagrangian space correlation function of haloes that have the 
same mass.  It is usual to use $S_n$ to denote the corresponding 
ratios of the mass correlation functions; for a Gaussian random 
field, $S_n=0$.  In the large volume limit, 
$\bar\xi_2 = (S_0/\delta_1^2)\,(1-2\nu^2)$, where 
$\nu$ is related to the halo mass (equation~\ref{pbsplit}).  
In this limit, equation~(\ref{muaqd0}) with the 
asymptotic expansion of the error function yields 
\begin{eqnarray}
H_3 &\to& {9\nu^2\,(\nu^2-1)\over (1-2\nu^2)^2} \nonumber \\
H_4 &\to& {4\nu^2\,(-3 + 24\nu^2 - 16\nu^4)\over (1-2\nu^2)^3}\nonumber \\
H_5 &\to& {125\nu^4\,(3 - 10\nu^2 + 5\nu^4)\over (1-2\nu^2)^4}.
\label{hnlag}
\end{eqnarray}
For massive haloes in this large cell limit  
\begin{equation}
H_n \to (n/2)^{n-1}\qquad\ {\rm when}\ \ \nu\gg 1.  
\label{hnmassive}
\end{equation}
These values are smaller than those associated with high peaks in a 
Gaussian random field, for which $H_n=n^{n-2}$.  This is a 
consequence of volume exclusion.  (For volume exclusion effects 
associated with peaks, see Coles 1986 and 
Lumsden, Heavens \& Peacock 1989.)  

It is interesting that these values are just those associated with 
the Poisson limit of the Generalized Poisson distribution 
(see, e.g., Saslaw \& Sheth 1993).  
Thus, equation~(\ref{hnmassive}) shows that, when smoothed on 
large scales, the Lagrangian space distribution of massive haloes 
is Poisson.  

\section{Generic Gaussian initial conditions}\label{scfics}
Section~\ref{wnics} provided expressions for the constrained and 
unconstrained halo mass functions, and for the moments of the 
halo counts in cells distribution, for the special case of 
white noise initial conditions.  
It is known that for more general Gaussian 
initial conditions [i.e., the initial power spectrum differs 
from $P(k)\propto k^0]$, the constrained and unconstrained mass 
functions have the same form as the white noise functions, 
provided that all quantities are written in terms of the variance, 
defined by equation~(\ref{svar}).  
That is, the unconditional and conditional mass functions for 
different initial power spectra differ only because the 
transformation from variance to mass depends on the initial power 
spectrum.  For example, if $P(k) \propto k^n$, then 
$S(M)\propto R^{-(n+3)}\propto M^{-(n+3)/3}$, where we have 
used the additional fact that in Lagrangian space $M$ and $V$ 
are equivalent variables.  Recall that white noise has $n=0$, 
so in the previous section $S\propto 1/M$.  

This section assumes that what works for the mass functions works 
for the counts in cells distributions also.  That is, expressions 
for the moments of halo distribution, when written in terms of 
the variance, are assumed to have the same form for all power 
spectra.  There is no compelling reason why this should be so.  
For example, the form of equation~(\ref{mualpha}) follows from 
the mutual independence of disconnected subvolumes.  While this 
is a reasonable assumption for white noise initial conditions, 
it is almost certainly wrong for other power spectra.  
Nevertheless, the hope is that those correlations between 
neighbouring volumes which are ignored when using 
equation~(\ref{mualpha}) to estimate halo--halo correlations will 
not make a crucial difference to the final answer, for reasons 
discussed by Bower (1991).  Moreover, Sheth \& Lemson (1998) 
showed that this simple model for the higher order moments 
associated with the forest of merger history trees is in reasonably 
good agreement with the results of numerical simulations, even 
when the initial distribution is quite different from white noise.  
Since it is these same higher order moments that one uses to 
estimate halo--halo correlations, their results suggest that this 
simple model should be reasonably accurate here as well.  

Another way to see why this conjecture should be accurate is 
the following.  The correlation function of haloes of two different 
masses is the product of the mean number of haloes of each of the 
two mass ranges times one plus the halo-halo correlation function.  
In principle, all three terms depend on power spectrum, although 
we only know this dependence for the two mean terms, and not for the 
correlation function.  In the white-noise case, were it not for 
volume exclusion, this correlation term would be zero.  For other 
initial power spectra, our conjecture means that we adjust the two 
mean terms correctly, and assume that most of the contribution to 
the correlation term comes from volume exclusion effects.  This 
means that our conjecture does correctly account for some, if not 
most, of the dependence of the correlation function and other 
higher order moments on the initial power spectrum.  

The integral (equation~\ref{c120q}) for the cross correlation 
between haloes of mass $M_1$ and $M_2$, that one obtains by 
ignoring these correlations can be solved analytically.  The 
final expression is lengthy, so we have not written it out 
below.  In the limit of large cells, i.e., $M_0\gg(M_1+M_2)$, 
\begin{equation}
\bar\xi_{\rm hh}^{\rm L}(12|0)\to B(1|0)\,B(2|0)\,S_0\ + 
\ {\rm correction\ terms},
\end{equation}
provided $\delta_1>\delta_0$.  In general, 
the correction terms are not as simple as in the white noise case, 
so we have not written them down explicitly.  

\begin{figure}
\centering
\mbox{\psfig{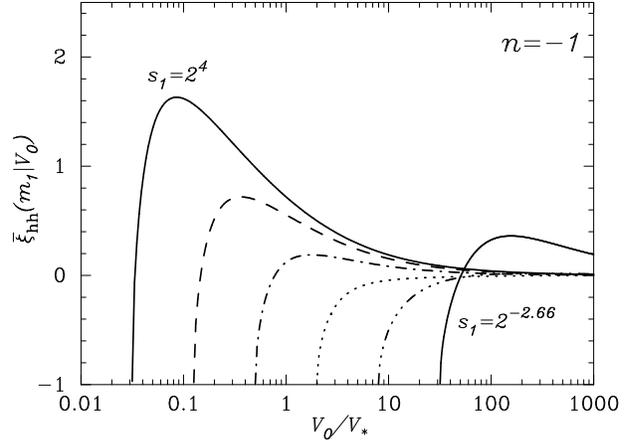}}
\caption{The volume average of the halo-halo correlation function, 
$\bar\xi_{\rm hh}^{\rm L}(11|0)$, given by equation~(\ref{c120q}), 
as a function of cell size $V_0$, when the initial power spectrum 
has slope $n=-1$.  
The different curves are for haloes with mass $m_1=s_1^{-3/2}$, 
and $s_1=2^4, 2^{2.66}, 2^{1.33}, 1, 2^{-1.33}$ and $2^{-2.66}$, 
respectively. }
\label{xihh1}
\end{figure}
\begin{figure}
\centering
\mbox{\psfig{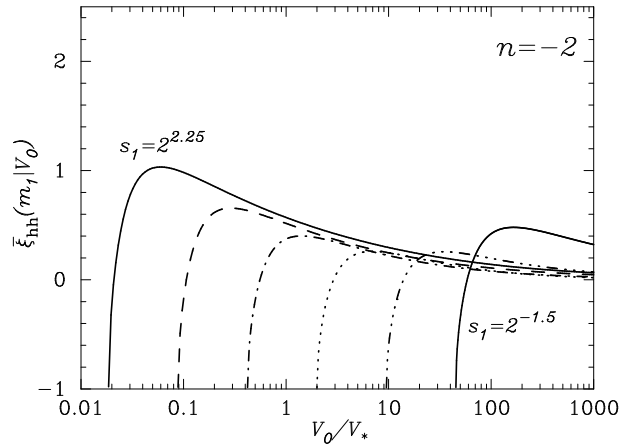}}
\caption{The volume average of the halo-halo correlation function, 
$\bar\xi_{\rm hh}^{\rm L}(11|0)$, given by equation~(\ref{c120q}), 
as a function of cell size $V_0$, when the initial power spectrum 
has slope $n=-2$.  
The different curves are for haloes with mass $m_1=s_1^{-3}$, 
and $s_1=2^{2.25}, 2^{1.5}, 2^{0.75}, 1, 2^{-0.75}$, and $2^{-1.5}$, 
respectively. }
\label{xihh2}
\end{figure}

In general, the full expression for halo--halo correlations 
differs from the white-noise expression in three significant ways.  
Firstly, ${\cal N}(2|10){\cal N}(1|0) = {\cal N}(1|20){\cal N}(2|0)$ 
only when $M_0$ is much greater than either $M_1$ or $M_2$.  This 
implies that, in general, equation~(\ref{fcross}) should be replaced 
with either $c(2|10) = {\cal N}(2|10){\cal N}(1|0)$, or $c(1|20)$, 
where ${\cal N}(2|10)$ is understood as the average number of 
$M_2$ haloes within those $M_0$ haloes that are known to have an 
$M_1$ halo in their central volume element.  The lack of spatial 
correlations for a white-noise spectrum meant that there, this 
restriction was irrelevant.  Here, however, this means that 
$\bar\xi_{\rm hh}^{\rm L}(12|0)$ and $\bar\xi_{\rm hh}^{\rm L}(21|0)$ 
computed using equation~(\ref{c120q}) are no longer equivalent.  

Secondly, the halo-halo correlations depend on the masses of the 
haloes themselves, rather than just their sum.  This suggests that 
volume exclusion effects are not the sole cause of halo 
correlations.
Thirdly, provided $S$ varies as some inverse power of scale, then,
in the limit of large separations, sufficiently high mass haloes 
are more correlated than low mass haloes.  
The correlation function of peaks in Gaussian random fields is 
known to depend exponentially on peak height 
(e.g., Bardeen et al. 1986; Jensen \& Szalay 1986; 
Lumsden, Heavens \& Peacock 1989; Reg\"os \& Szalay 1995).  
If high mass haloes correspond to high peaks in the initial 
density field, then this result is qualitatively similar to 
that for peaks.  The agreement with the peaks results is only 
qualitative.  For example, just as in the white-noise case, 
the higher-order moments of the spatial distribution of massive 
haloes is different from that of high peaks.  

Figures~\ref{xihh1} and~\ref{xihh2} show the volume average of 
the halo-halo correlation function (equation~\ref{c120q}) as a 
function of scale, when the initial power spectrum has slope 
$n=-1$ and $n=-2$, respectively.  
A range of choices of halo mass are shown.  
On scales smaller than $2\,v_1$, volume exclusion effects mean 
that $\bar\xi_{\rm hh}^{\rm L} = -1$.  As a result of halo 
exclusion effects, haloes less massive than $M_*$ are positively 
correlated on intermediate scales, and on scales larger than about 
$4\,v_1$, $\bar\xi_{\rm hh}^{\rm L}(11|0)/s_0\approx$ constant.  
On sufficiently large scales, haloes that are more massive than 
$\sim M_*$ are  more correlated than less massive haloes.  

\section{The halo distribution in Eulerian space}\label{scmow} 
The previous sections showed how to quantify the difference 
between the halo and matter distributions in Lagrangian space.  
Dynamical evolution changes these distributions, so the bias 
between haloes and mass in Eulerian space is likely to be different 
from that initially.  

Mo \& White (1996) argued that the bias relation in Eulerian space, 
i.e., the mean overdensity of $\delta_1$-haloes that are in spheres 
with comoving volume $V$ which contain mass $M_0$ at $z$, so they 
have Eulerian overdensity 
\begin{equation}
\Delta\equiv 1+\delta\equiv  M_0/\bar\rho V, 
\label{deleul}
\end{equation}
should be 
\begin{equation}
\delta_{\rm h}^{\rm E}(1|0) = 
{{\cal N}(M_1,\delta_1|M_0,\delta_0)\over \bar n(M_1,\delta_1)V} - 1,
\end{equation}
where ${\cal N}(1|0)$ is given by the (Lagrangian) 
equation~(\ref{n10}), but with 
\begin{equation}
{\delta_0\over 1+z} = 1.686 - {1.35\over\Delta^{2/3}} 
+ {0.788\over\Delta^{0.587}} - {1.124\over\Delta^{1/2}}.
\label{d0de}
\end{equation}
Therefore, in their model, expressions for the higher order 
moments of the bias relation in the Eulerian space can be 
obtained by transforming the corresponding Lagrangian 
expressions similarly.  We will use this fact below.  

Let $p(M_0|V,z)\,{\rm d}M_0$ denote the probability that an 
Eulerian cell $V$ contains mass in the range 
${\rm d}M_0$ of $M_0$ at $z$.  We will sometimes call this the 
Eulerian probability distribution function.  Of course, 
$p(M_0|V,z)\,{\rm d}M_0 = p(\Delta|V,z)\,{\rm d}\Delta$ and 
\begin{equation}
\int_0^\infty p(\Delta|V,z)\,{\rm d}\Delta = 
\int_0^\infty \Delta\ p(\Delta|V,z)\,{\rm d}\Delta = 1.  
\label{pdfnorm}
\end{equation}
Let ${\bar N}(M_1,\delta_1|M_0,V,z)$ 
denote the average number of $(M_1,\delta_1)$-haloes in 
such a cell.  Then the average number of haloes in Eulerian 
cells of size $V$ is 
\begin{equation}
n(M_1,\delta_1)V \equiv 
\int_0^\infty {\bar N}(M_1,\delta_1|M_0,V)\ p(M_0|V)\,{\rm d}M_0 ,
\label{meaneul}
\end{equation}
where we have not bothered to write the dependence on $z$ 
explicitly.  This is the analogue of the Lagrangian 
relation~(\ref{feq10}).  Suppose we assume that 
\begin{equation}
{\bar N}(M_1,\delta_1|M_0,V,z) = {\cal N}(M_1,\delta_1|M_0,\delta_0), 
\label{equalns}
\end{equation}
where $\delta_0$ is given by equation~(\ref{d0de}).  
That is, the average number of haloes in Eulerian cells of size $V$ 
that contain mass $M_0$ is assumed to be the same as the average 
number of haloes in Lagrangian cells $M_0$ that, because they 
originally had overdensity $\delta_0(\Delta)$, they have 
size $V$ at $z$.  Then equation~(\ref{meaneul}) implies that
\begin{equation}
f(M_1,\delta_1) = 
\int_{M_1}^\infty f(M_1,\delta_1|M_0,\delta_0)\ 
\Delta\ p(M_0|V)\,{\rm d}M_0 ,
\label{volt}
\end{equation}
where $\delta_0$ is given by equation~(\ref{d0de}), and again, we 
have not written the $z$ dependence explicitly.  
The lower limit of the integral has been set to $M_1$ since, in 
the spherical collapse model which gives equation~(\ref{d0de}), 
the Eulerian radius of a collapsed halo is zero.  This means that 
if an Eulerian cell $V$ contains an $(M_1,\delta_1)$-halo, then 
it must contain all of the halo's mass, so it must have 
$M_0\ge M_1$.  

Equation~(\ref{volt}) is interesting for the following reason.  
The term on the left hand side, $f(M_1,\delta_1)$ is known.  
If the Eulerian cell size $V$ is given, then 
$f(M_1,\delta_1|M_0,\delta_0)$ is also known, for all $M_0$.  
Only the Eulerian probability distribution $p(M_0|V)$ is not 
known.  Therefore, equation~(\ref{volt}) is an integral equation 
of the first kind, so it can be solved numerically to yield 
$p(M_0|V)\,{\rm d}M_0$.  

That is, for any Eulerian cell size $V$, the 
assumption~(\ref{equalns}) allows one to solve for the 
Eulerian probability distribution function that is associated 
with the spherical collapse model as parametrized by 
equation~(\ref{d0de}).  Once $p(M_0|V)$ is known, repeated use 
of the assumption~(\ref{equalns}) allows one to compute 
\begin{eqnarray}
\bar\xi_{\rm hm}^{\rm E}(M_1,\delta_1|V) &\equiv& 
\Bigl\langle\delta_{\rm h}^{\rm E}(1|0)\,\delta\Bigr\rangle \nonumber \\
&=& \int_0^\infty \delta_{\rm h}^{\rm E}(1|0)\,\delta\ 
p(M_0|V)\,{\rm d}M_0,
\label{xihme}
\end{eqnarray}
where $(1+\delta)\equiv M_0/\bar\rho V$.  
Notice that this resembles the Lagrangian relation~(\ref{xihml10}).  
Similarly, 
\begin{eqnarray}
&&\!\!\!\!\!\!\!\!\!\!\!\!
1+\bar\xi_{\rm hh}^{\rm E}(M_1,M_2,\delta_1|V) = \nonumber \\
&&\int_0^\infty {c(M_1,M_2,\delta_1|M_0,\delta_0)\over 
n(M_1,\delta_1)V\,n(M_2,\delta_1)V}\ p(M_0|V)\,{\rm d}M_0 ,
\label{xihhe}
\end{eqnarray}
where $c(12|0)={\cal N}(1|0){\cal N}(2|10)$ is the Lagrangian 
relation~(\ref{fcross}), with $\delta_0$ given by 
equation~(\ref{d0de}).  

Our approach extends that of Mo \& White (1996).  
They wrote down equations~(\ref{xihme}) and~(\ref{xihhe}), 
though they did not have an expression for $c(12|0)$.  However, 
they did not write down equation~(\ref{meaneul}), so they 
did not know how to solve for the Eulerian $p(M_0|V)$.  
Therefore, they assumed that they could use the one measured in 
their simulations.  Strictly speaking, this is not permitted, since 
there is no guarantee that then equation~(\ref{meaneul}) is 
satisfied, as it should.  Indeed, if one substitutes the Lognormal 
distribution for $p(M_0|V)$ (as Mo \& White did) into this formula, 
then one finds that, in general, this normalization requirement is 
not satisfied (though Mo \& White do not mention this).  
Nevertheless, if the spherical model is a good approximation to 
what actually happens in the simulations, then there is some hope 
that using the actual $p(M_0|V)$ distribution measured in the 
simulations will, indeed, give the correct normalization.  
(Also see Sheth 1998 for more discussion of this point.)  

Below, when we compare our results with simulations, we will 
show that the Mo \& White approach is reasonably well normalized 
on large scales.  So, although we should first determine the 
Eulerian $p(M_0|V)$ using the integral equation~(\ref{volt}), 
and then we should use it to compute $\bar\xi_{\rm hm}^{\rm E}$ 
and $\bar\xi_{\rm hh}^{\rm E}$ self-consistently, 
in what follows, we will not.  

Mo \& White mainly considered the case in which the time at 
which the haloes first virialized $a_1$, and that when their 
spatial distribution was studied $a_0$, were the same.  
They also studied the spatial distribution of haloes at epochs 
later than those at which the haloes had virialized ($a_0\ge a_1$).  
In both these cases, the previous formulae are correct if 
$\delta_1=1.68647\,(a_0/a_1)$ and $\delta_0$ is given by 
equation~(\ref{d0de}) with $z=0$.  Thus, $\delta_1\ge\delta_0$ is 
always satisfied.  

In principle, it should also be possible to use the spherical 
model to describe the distribution of the haloes at high redshift, 
prior to virialization.  
The spatial distribution at some early time $z_{\rm i}$ 
of haloes that will virialize at the present $z=0$, 
is described by the previous expressions, but with the appropriate 
value of $z=z_{\rm i}$ in equation~(\ref{d0de}).  
This means that we need to know the Eulerian distribution function 
as a function of $z$.  For example, for haloes that virialize 
at $z=0$, $\delta_1=1.68647$, so it is possible that 
$\delta_0>\delta_1$.  In the language of the previous sections, 
such Eulerian cells are not isolated.  So, in principle, we 
need to be able to compute the probability that an Eulerian cell 
is isolated.  In general, this is difficult.  
Fortunately, things simplify when $z\gg 1$: in this limit 
$S\ll 1$, most fluctuations are small $(|\delta|\ll 1)$, and 
the Eulerian distribution function tends to a Gaussian.  
So, in this limit, this procedure reduces to the Lagrangian 
description of the previous sections.  It is also reassuring that, 
in this limit, the spherical model expressions reduce to those 
expected using linear theory (Section 19 in Peebles 1980).  
We will use this fact below.  

\section{Comparison with simulations}\label{sims}
This section shows the results of comparing the model predictions 
obtained in the previous sections to the halo distributions 
measured in numerical simulations of clustering.  
This is done in two steps.  First, the theoretical bias relation, 
$\delta_{\rm h}(1|0)$, and the scatter around this 
relation, are compared with those found in the simulations.  
Then the theoretical halo--mass and halo--halo correlation 
functions are compared with those in the simulations, since 
these are essentially weighted integrals over the bias relations.  
We do this in Lagrangian space, and then in Eulerian space.  

The simulations used here are the same as those used by 
Mo \& White (1996), where they are described in more detail.  
They follow the evolution of $10^6$ identical particles 
in a cubic box with periodic boundary conditions.  If the volume 
$L^3$ of the box, the mass $m$ per particle, and the initial 
expansion factor $a$ are all set to unity, then the 
simulations are normalized so that $S(M) = M^{-(n+3)/3}$ 
initially, where $n$ is the initial slope of the power spectrum.  
The characteristic mass $M_*(a)$ at the expansion time $a$ is 
given by $S(M_*) = (\delta_{\rm c}/a)^2$, for some $\delta_{\rm c}$ 
which is determined by fitting the unconditional mass function of 
equation~(\ref{nm1}) to the mass function of bound objects 
identified in the simulations.  
The group identification algorithm used here is the same 
friends-of-friends algorithm used by Mo \& White, as are the 
methods for assigning Lagrangian and Eulerian positions to a 
group identified at any given time.  
As for the simulations studied by Lacey \& Cole (1994), 
the mass function of bound objects in these simulations is fit, 
to within a factor of two or so, by equation~(\ref{nm1}) with 
$\delta_{\rm c}=1.7$.  This value is used to compute all the 
theoretical curves shown below.  
  
The main complication in comparing the theory to simulations 
is that of the finite mass resolution in the simulations.  
This means that, in practice, correlations between haloes 
are measured for a range of masses.  This has an important 
consequence, since now, the distribution of isolated regions 
is different from that of the centre-of-mass distribution of 
collapsed haloes (this is a subtle point that is discussed 
more fully in Section~\ref{range}).  This is unfortunate, 
since, to account for this fact, we must make some assumption 
about the nature of the Lagrangian space volume elements associated 
with halo centres-of-mass.  In the Poisson and white noise 
cases, Section~\ref{range} argued that we could simply assume that 
this volume element is just a randomly chosen one of the volume 
elements of a halo.  This assumption is almost certainly wrong 
if the initial distribution differs from white noise.  
Nevertheless, for reasons discussed in Section~\ref{range}, 
we will assume that this is, indeed, the case.  

This means that the mean bias relation is 
\begin{equation}
\delta_{\rm h}(>\!m|0) = 
{{\cal N}(>\!m,\delta_1|M_0,\delta_0)\over n(>\!m,\delta_1)V_0} - 1,
\end{equation}
where 
\begin{displaymath}
n(>\!m,\delta_1) = \int_m^\infty n(M_1,\delta_1)\ {\rm d}M_1,
\end{displaymath}
and
\begin{displaymath}
{\cal N}(>\!m,\delta_1|M_0,\delta_0) = 
\int_m^\infty {\cal N}(M_1,\delta_1|M_0,\delta_0)\ {\rm d}M_1,
\end{displaymath}
provided $M_1\le M_0$ and $\delta_1\ge \delta_0$.   
In the Eulerian space, $M_0$ and $\delta_0$ are obtained from $V$ 
and $\delta$ as described in Section~\ref{scmow}.  

This quantity depends only on the first moment of the subclump 
distribution.  Although it could have been computed by 
Mo \& White (1996), they did not show it.  
The scatter in this relation depends on the second order moment, 
so, although they were unable to compute it, we can.  

\begin{figure}
\centering
\mbox{\psfig{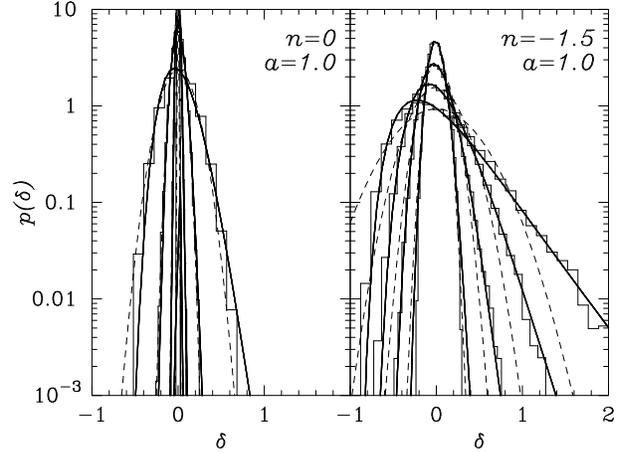}}
\caption{The Lagrangian space probability distribution function 
$p(\delta)$ as a function of overdensity $\delta$.  
Each panel shows four choices of scale 
$R/L = 0.02$ (broadest curves), $0.4$, $0.8$ and $0.16$ 
(narrowest curves).  
Histograms show the distribution measured in the simulations; 
thin dashed curves show Gaussian distributions, and thicker 
solid curves show Generalized Inverse Gaussian fitting functions 
(equation~\ref{ginvg}), that have the same variance.}
\label{lagpdf}
\end{figure}

There are additional reasons why it is not entirely straightforward 
to compare the theory with simulations.  For example, the 
average number density of haloes (the unconditional mass function), 
and the average number of subhaloes within haloes 
(the conditional mass function) in the simulations are, typically, 
described by the theory only to within a factor of two or so.  
Also, on small scales in particular, the initial particle 
distribution in the simulations is not particularly Gaussian 
when the initial power on large scales is significant 
(see Fig.~\ref{lagpdf}).  
Since the bias relations are essentially the ratio of the conditional 
to the unconditional mass functions, they are sensitive to the 
first of these discrepancies.  The integrals which define 
$\bar\xi_{\rm hm}^{\rm L}$ and $\bar\xi_{\rm hh}^{\rm L}$ are 
also sensitive to the shape of the initial probability distribution 
function, so they are sensitive to both these discrepancies.  

Finally, there is some uncertainly regarding how the initial particle 
load in the simulations should be treated.  This freedom arises 
because the initial particle distribution is not the true Lagrangian 
distribution, but a linearly evolved version of it.  This means 
that, when comparing the Lagrangian theory with the simulations, 
we must account for the fact that cell sizes in the initial 
distribution are not the same as the associated Lagrangian size.  
Though they do not say so in their paper, Mo \& White (1996) 
treated this problem as follows (private communication).  
They used equation~(\ref{deleul}) to rescale the size of each cell 
in the simulations, and then used this rescaled size in the 
denominator that defines $\delta_{\rm h}^{\rm L}$, but nowhere 
else.  They then used this value of $\delta_{\rm h}^{\rm L}$ 
when averaging over all cells to determine what they called the 
Lagrangian $\bar\xi_{\rm hm}^{\rm L}$.  

We have chosen the following procedure.  
We treat the initial particle distribution no differently from any 
other output time in the simulations.  
This means that we plot the simulation results exactly as 
measured, with no rescaling.  We then compare these to our 
theoretical Eulerian expressions, transformed according to the 
spherical model to the appropriate redshift.  Recall that, 
in the limit of small initial fluctuations, this is the same as 
using linear theory to make the necessary corrections 
(Section~19 of Peebles 1980).
The complication is that, in this case, the associated 
$p(\delta)$ distribution is no longer Gaussian, so the 
distribution corresponding to $q$ is no longer known.  
Nevertheless, if $p(\delta)$ is sufficiently close to Gaussian, 
then using $q$ should be a good approximation.  We find that 
the Generalized Inverse Gaussian distributions (described in 
Section~\ref{gig} below) provide reasonable fits to the counts 
in cells distributions measured in the simulations for a wide 
range of scales and output times, so we use them for $p(\delta)$.  

\subsection{Biasing in Lagrangian space}
This subsection compares the bias relation between haloes and 
mass measured in the simulations in the Lagrangian space with 
the theoretical model developed in the previous sections.  

\begin{figure}
\centering
\mbox{\psfig{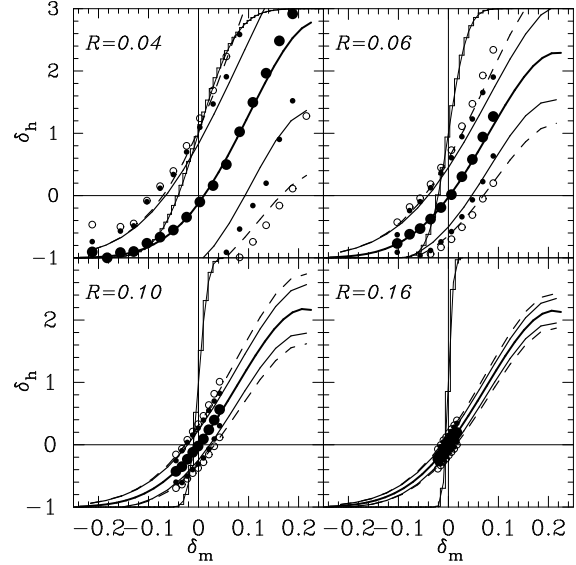}}
\caption{The Lagrangian space bias relation for haloes which contain 
more than $m=32$ particles that form from white noise initial 
conditions.  Plot shows the mean overdensity of haloes 
$\delta_{\rm h}(>\!m|V)$ as a function of the overdensity of 
mass $\delta_{\rm m}$ in spherical cells of radius $R$, as 
well as the scatter around the mean.  
Symbols show quantities measured in the simulations: 
large filled circles show the mean, 
smaller filled circles show the rms scatter, 
and open circles show the scatter if the halo counts were Poisson.  
Solid curves show the model predictions, dashed curves show the 
Poisson scatter corresponding to the theoretical mean.  
Haloes were identified at an expansion factor of $a=6.1$; 
the bias relation was computed from the halo-centre-of-mass 
and mass distributions at the initial time $a=1$.  The histograms 
that rise from left to right in each panel show the cumulative 
counts-in-cells distribution.  The simulations provide a good 
test of the theory only in the range where this cumulative curve 
is steep.}
\label{bias0l3}
\end{figure}

\begin{figure}
\centering
\mbox{\psfig{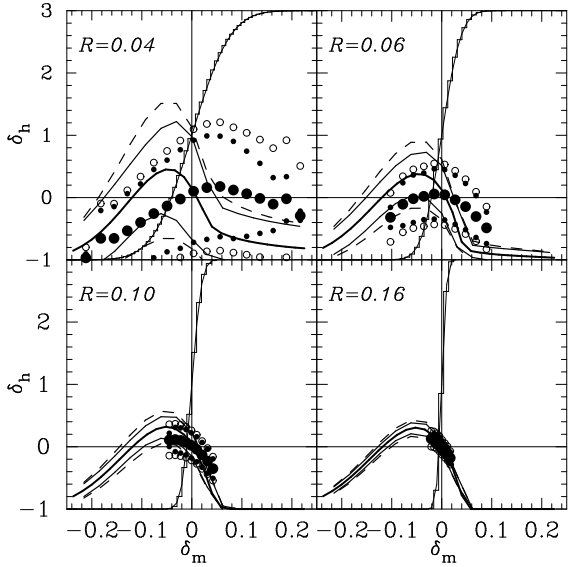}}
\caption{The same as the previous Figure, i.e., $n=0$ and $m=32$, 
but now $a=36.9$.  }
\label{bias0l732}
\end{figure}

\begin{figure}
\centering
\mbox{\psfig{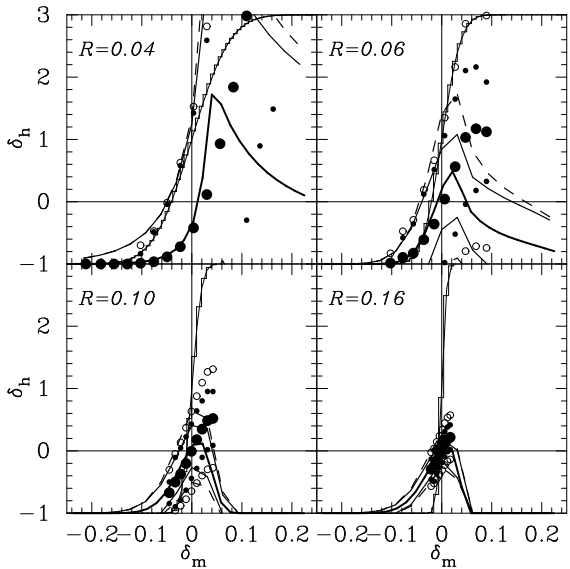}}
\caption{The same as the previous Figure, i.e., $n=0$ and $a=36.9$, 
but now $m=256$.  }
\label{bias0l7256}
\end{figure}

Figs.~\ref{bias0l3}---\ref{bias1l7256} show the bias relation for 
haloes containing more than $m$ particles, identified in 
simulations with initial power spectra having slope $n$ at 
an expansion factor $a$ since the initial time, and for four 
representative choices of the spherical cell radius:  
$R/L = 0.02$, $0.04$, $0.08$, and $0.16$.  
For each cell size, statistics were averaged over 27,000 spherical 
cells.  The histogram which rises from the bottom left to 
the top right of each panel shows the cumulative distribution 
function of the matter fluctuation $\delta_{\rm m}$.  
This curve is intended to show the range of $\delta_{\rm m}$ over 
which the simulations are able to provide a good test of the theory.  
The thin dashed line through each histogram shows the corresponding 
cumulative distribution for a Gaussian with the same variance; the 
thin solid line through each histogram shows the corresponding 
Generalized Inverse Gaussian.  
The large filled circles show the mean bias relation measured in the 
simulations, smaller filled circles show the rms fluctuations around 
this mean, and the open circles show the expected Poisson 
fluctuation given the mean.  In most cases, the rms fluctuations 
are smaller than the Poisson value; this shows that volume 
exclusion effects are important.  The thickest solid curve shows the 
mean bias relation predicted by the model, the less thick solid curves 
show the theoretical rms fluctuation around this mean, and the 
dashed curves show the value if the fluctuations were Poisson.  

Fig.~\ref{bias0l3} is extremely encouraging.  
The theory is able to describe the mean bias relation, 
$\langle\delta_{\rm h}|\delta_{\rm m}\rangle$, 
as well as the scatter in this relation well, 
even when the scatter is less than Poisson (though, for haloes 
of this mass range, the difference from Poisson scatter is small).  
That is, the theory appears to describe the effects of volume 
exclusion on the halo distribution well.  
Figs.~\ref{bias0l732} and~\ref{bias0l7256} are intended to show 
that the theory must be used with some caution.  These figures 
show the bias relation associated with haloes identified at a 
later time than those in Fig.~\ref{bias0l3}.  
Since $M_*\equiv (a/\delta_{\rm c})^2$ for white noise, 
$M_*\approx 13$ for the haloes in Fig.~\ref{bias0l3}, 
whereas $M_*\approx 470$ for the haloes in 
Figs.~\ref{bias0l732} and~\ref{bias0l7256}.  
At the later output time, the theory gets the mean of the 
Lagrangian bias relation wrong, although the scatter around the 
mean is still qualitatively correct, when the minimum mass $m=32$.
For haloes more massive than $m=256$, however, the theory is 
accurate, in the mean, and for the scatter.  

\begin{figure}
\centering
\mbox{\psfig{figure=biasf8.ps,height=7.5cm,bbllx=94pt,bblly=5pt,bburx=618pt,bbury=515pt}}
\caption{The same as the previous Figure, but for $n=-1.5$, $m=32$ 
and $a=6.07$.  }
\label{bias1l732}
\end{figure}
\begin{figure}
\centering
\mbox{\psfig{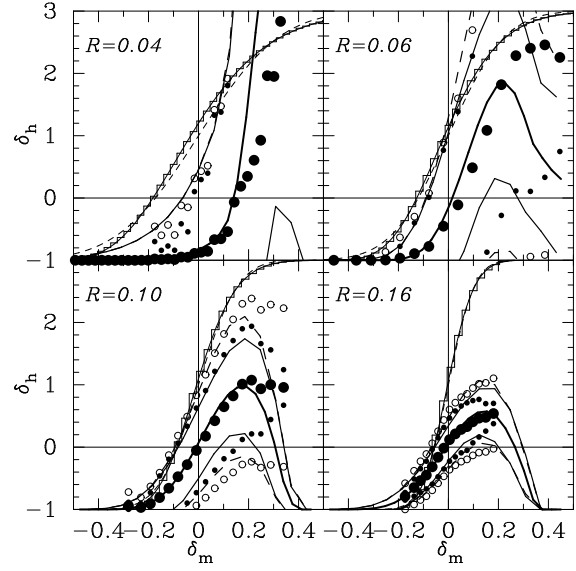}}
\caption{The same as the previous Figure, i.e., $n=-1.5$ and $a=6.07$ 
but now $m=256$.  }
\label{bias1l7256}
\end{figure}

Thus, these figures show that the theory is relatively accurate 
when describing the distribution of haloes more massive than 
$\sim M_*$, but not of less massive haloes.  This suggests that 
the spherical model is a good description of the collapse of 
massive haloes, but that the formation and evolution of less 
massive haloes may be more complicated.  

Figs.~\ref{bias1l732} and~\ref{bias1l7256} show that the theory 
works even when the initial conditions are different from 
white noise.  These figures were constructed from haloes identified 
at an expansion factor $a=6.1$ in a simulation in which the initial 
power spectrum had slope $n=-1.5$.  So, for these figures, 
$M_*\approx 163$.  Again, the bias relation associated with 
massive haloes is well described by the theory 
(Fig.~\ref{bias1l7256}), whereas that of the less massive haloes 
is not (Fig.~\ref{bias1l732}).  

Before concluding this subsection, it is worth noting that the 
theoretical curves for the mean bias relation become increasingly 
different from the simulation results as $R$ decreases.  
Although the mean relation on these smaller scales is different, 
the predicted scatter around the mean shows the same qualitative 
behaviour as that measured.  We have not shown curves for smaller 
$R$ here, since on these smaller scales it is not clear how much 
of the discrepancy in the mean is due to limitations associated 
with the finite number of particles in the numerical simulations.  

\subsection{Lagrangian space halo correlation functions}

\begin{figure}
\centering
\mbox{\psfig{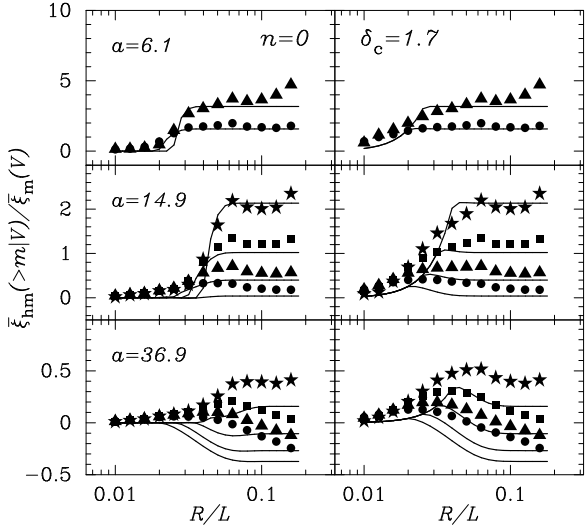}}
\caption{The volume average of the Lagrangian space halo-mass 
cross correlation function, $\bar\xi_{\rm hm}(>\!m|R)$, 
equation~(\ref{cumhm}), as a function of cell size $R$, 
when the initial power spectrum has slope $n$, for haloes identified 
at a range of output times, labelled by the expansion factor $a$.  
Panels on the left show the result of computing the average 
by using only those cells whose initial density was less  
than $\delta_{\rm c}/a$.  Panels on the right show the result 
of averaging over all Lagrangian cells, whatever their density.  
Symbols show this quantity measured in the simulations; 
curves show the model predictions (made using the value of 
$\delta_{\rm c}$ shown).  
From bottom to top in each panel, the different curves are for 
haloes with $m=32$ (circles), 64 (triangles), 128 (squares), 
and $256$ (stars) particles.}
\label{simhm0l}
\end{figure}

The cross correlation between haloes and mass is essentially 
a weighted integral over the bias relations shown in the previous 
subsection.  In this sense, $\bar\xi_{\rm hm}$ is a slightly less 
fundamental quantity than 
$\langle\delta_{\rm h}|\delta_{\rm m}\rangle$.  
The cross correlation between haloes with mass larger than $m$, 
whose centres-of-mass are within a cell $V_0$, and the mass within 
that cell, is 
\begin{eqnarray}
\bar\xi_{\rm hm}(>\!m|0) &=& 
\int_m^\infty {n(M_1,\delta_1)\over n(>\!m,\delta_1)}\ 
\bar\xi_{\rm hm}(1|0)\ {\rm d}M_1 \nonumber \\
&&\ + \ \ \delta_1 \
\int_\mu^\infty 
{n(M_1,\delta_1)\over n(>\!m,\delta_1)}\ {\rm d}M_1
\label{cumhm}
\end{eqnarray}
where $n(M_1,\delta_1)$ and $\bar\xi_{\rm hm}(1|0)$ were 
defined earlier, $\mu={\rm max}(m,M_0)$, 
and the convention is that, in the first term, 
$\bar\xi_{\rm hm}(1|0) = 0$ if $M_1>M_0$.  
The second term accounts for the difference between counting  
haloes instead of isolated regions.  This is the analogue of 
equation~(\ref{rangehm}).  In general, these integrals over the 
range of halo masses must be done numerically.  

Figure~\ref{simhm0l} shows equation~(\ref{cumhm}) for white noise 
initial conditions, for haloes identified at a range of output 
times, and minimum mass cutoffs, as a function of scale.  
The plots are for the Lagrangian space distribution of haloes 
identified at the epoch $a$, and the four curves in each plot 
are (from bottom to top) for $m=32, 64, 128$, and $256$ particles, 
respectively.  
The figure actually shows $\bar\xi_{\rm hm}/\bar\xi_{\rm m}$, 
where $\bar\xi_{\rm m} \equiv a^2/M^\alpha$, and 
$S_*(a)\equiv \delta_{\rm c}^2$, with $\delta_{\rm c}=1.7$ 
as required by the spherical model.  
The two panels show the difference between averaging over all 
Lagrangian cells (right) and averaging only over those Lagrangian 
cells which are not too overdense (left).  
Thus, the panels on the left are the same quantity computed by 
Mo \& White (1996).  

\begin{figure}
\centering
\mbox{\psfig{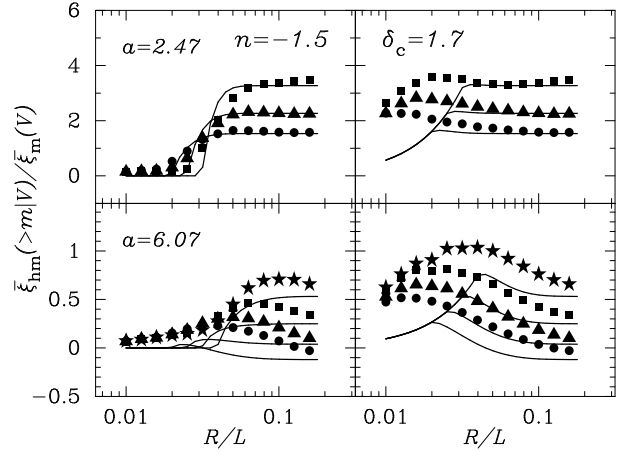}}
\caption{The same as the previous figure, but now $n=-1.5$.
The theory describes the simulation results reasonably well for 
massive haloes, and rather poorly for less massive haloes, 
where massive and less massive are defined relative to $M_*(a)$.}
\label{simhm1hl}
\end{figure}

Typically, the fits in the panels on the left are better than those 
shown in the panels on the right, and, typically, the fit is usually 
better on larger than on smaller scales.  (On large scales, the 
number of cells in the two panels is almost the same anyway.)
This suggests that the way in which the model assigns haloes 
to Lagrangian cells that are not isolated is not quite correct.  
In the panels on the right, the model systematically underestimates 
$\bar\xi_{\rm hm}(>m|V)$ on small scales.  Comparison with 
Fig.~\ref{simhm1hl} shows that the discrepancy increases as the 
initial power on large scales increases ($n$ becomes more negative).  
This is not unexpected.  The assumption that the centre-of-mass 
particle is a random one of a halo's particles is likely to 
be less accurate as $n$ becomes more negative.  
On the other hand, some of the discrepancy on small scales may 
be spurious.  These are measurements in Lagrangian space, 
and the initial inter-particle spacing was on the order of 
$R/L\sim 0.01$, so it is not clear that differences on these small 
scales are significant.  Moreover, recall that when $n=-1.5$, then 
the initial particle distribution on small scales is far from 
Gaussian (Fig.~\ref{lagpdf}).  

Figs.~\ref{simhm0l} and~\ref{simhm1hl} appear to show that the 
theory describes the simulation results better for small values 
of the expansion factor $a$.  This is a consequence of one of the 
results of the previous subsection; when the mass of a halo 
identified at time $a$ is expressed in units of $M_*(a)$, then 
the theory describes the distribution of massive haloes better than 
less massive ones.
At some small $a$, haloes with more than, say, 64 particles are 
larger relative to an $M_*$ halo at that time, than they are at 
some later time.  So, in Figs.~\ref{simhm0l} and~\ref{simhm1hl}, 
the theory appears to work better at small $a$ than large.  

Before considering the halo--halo correlation function we think it 
worth remarking that some of the agreement between theory and 
simulation is a consequence of showing the ratio 
$\bar\xi_{\rm hm}/\bar\xi_{\rm m}$, rather than 
$\bar\xi_{\rm hm}$ and $\bar\xi_{\rm m}$ themselves.  
On small scales $\bar\xi_{\rm m}\gg 1$, so the ratio tends to zero.  
Had we shown $\bar\xi_{\rm hm}$ only, then the theory and the 
simulation curves can look quite different, particularly on small 
scales.  Again, this suggests that the theory should be used with 
caution.  

The correlation function between haloes with mass greater than $m$, 
averaged over Lagrangian cells of size $V_0$, is 
\begin{eqnarray}
1+\bar\xi_{\rm hh}(>\!m|0)\!\!&=&\!\! 
\int_m^\infty\!\!\!{\rm d}M_1\int_m^\infty\!\!\!{\rm d}M_2\ 
{n(M_1,\delta_1)\,n(M_2,\delta_1)\over n^2(>\!m,\delta_1)}\nonumber\\
&&\qquad\qquad \times\ \  \Bigl[1+\bar\xi_{\rm hh}(12|0)\Bigr],
\label{cumhh}
\end{eqnarray}
where $\bar\xi_{\rm hh}(12|0)$ is given by equation~(\ref{c120q}), 
and the convention is that $\bar\xi_{\rm hh}(12|0)=-1$ if 
$M_1+M_2>M_0$.  This is the analogue of equation~(\ref{rangehh}).  

Figs.~\ref{xihh0}--\ref{xihh2} show that, as a result of 
volume exclusion effects, $\bar\xi_{\rm hh}(>\!m|0)$ is 
likely to be negative for all except large values of $V_0$.  
Since halo correlations increase as $n$ decreases, this effect 
will be weaker as $n$ becomes more negative.  Thus, when $n\sim 0$, 
then $\bar\xi_{\rm hh}(>m)$ will almost always be negative.  
Only when $n\sim -1$ or so will it become positive, and then, only 
when $m$ is large compared to $M_*(z)$.  The distribution of 
haloes measured in the simulations show that this is true.  

\begin{figure}
\centering
\mbox{\psfig{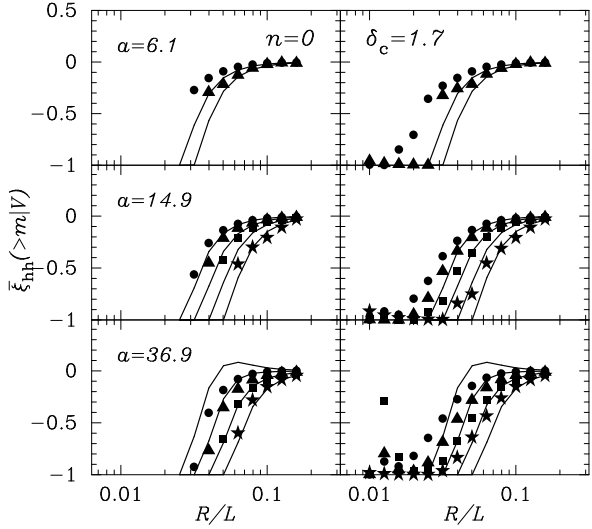}}
\caption{The volume average of the Lagrangian space halo-halo 
correlation function, $\bar\xi_{\rm hh}(>\!m|R)$, 
equation~(\ref{cumhh}), as a function of cell size $R$, 
for the same haloes that were used to make Fig.~\ref{simhm0l}.  
Panels on the left show the result of computing the average 
by using only those cells whose initial density was less  
than $\delta_{\rm c}/a$.  Panels on the right show the result 
of averaging over all Lagrangian cells, whatever their density.  
Symbols show this quantity measured in the simulations; 
curves show the model predictions with the value of $\delta_{\rm c}$ 
shown.}
\label{simhh0l}
\end{figure}

\begin{figure}
\centering
\mbox{\psfig{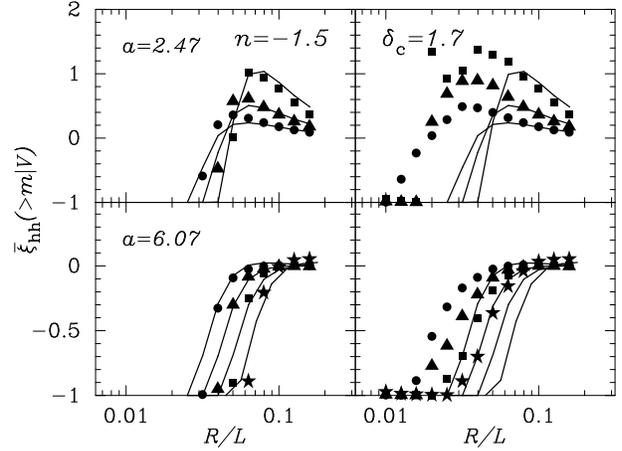}}
\caption{The same as the previous figure, but now $n=-1.5$.}
\label{simhh1hl}
\end{figure}

Fig.~\ref{simhh0l} shows equation~(\ref{cumhh}), 
for a range of output times and minimum mass cutoffs, 
as a function of scale.  The plots are for the Lagrangian space 
distribution of the same haloes that were used to produce 
Fig.~\ref{simhm0l}.  
Notice that more massive haloes are always less clustered 
than less massive haloes, in agreement with the white-noise result 
(equation~\ref{xilim}).  This would not have been expected from 
the Mo \& White (1996) formulae.  Again, this suggests that our 
model for halo exclusion effects is reasonably accurate.  
Figure~\ref{simhh1hl} shows that our model is also reasonably 
accurate when the initial conditions differ from white noise.  


There are, of course, some systematic differences.  
The theoretical curves fit the data in the panels on the left 
better than the data shown on the right, and the discrepancy 
is more obvious for $n=-1.5$ than for $n=0$.  This simply
reflects the fact that our model, in which the centre-of-mass 
particle of a halo is a random one of its constituent particles,
is not very realistic (though it is a better approximation in 
the white noise case).  Also, on small scales, the simulation haloes 
are systematically less anti-correlated than the model predictions, 
suggesting that they are affected less strongly by volume exclusion 
effects than in the model.  This is a consequence of at least two 
facts.  The first is that, in the simulations, small haloes in 
particular are not necessarily spherical, so the excluded volume 
associated with them is not necessarily spherical.  Thus, in the 
simulations, it is possible for two centre-of-mass particles, 
associated with haloes of mass $M_1$ and $M_2$, to fall in the same 
spherical Lagrangian region $M_0$, even if $M_1+M_2 > M_0$, since 
not all their associated particles actually fall in $M_0$.  
In the model this never happens.  The second is that, in fact, 
the number density of haloes described by the model 
(the denominator in equation~\ref{cumhh}) is, in general, only 
within a factor of two or so of the actual number density of haloes 
measured in the simulations.  Since the halo--halo correlation 
function is normalized by the square of this number density, this  
relatively minor discrepancy may still be important.  
Finally, recall that when $n=-1.5$, then the initial distribution 
on small scales was not particularly Gaussian (Fig.~\ref{lagpdf}).  

\subsection{The Eulerian probability distribution function}\label{gig}
We argued (Section~\ref{scmow}) that, in principle, the Mo \& White 
model for transforming Lagrangian space statistics into Eulerian 
space ones can be used to derive the Eulerian space dark matter 
distribution function.  
To do so, we showed that one must solve the integral 
equation~(\ref{volt}).  
However, not only must the resulting distribution be correctly 
normalized (to unity), but $\langle\Delta\rangle = 1$ as well
(cf. equation~\ref{pdfnorm}).  
There is no guarantee that, in general, the solution to the 
integral equation will meet both normalization conditions.  
Therefore, we have chosen to stick with the approach used by 
Mo \& White (1996).  Namely, when the Eulerian distribution 
function is required, we will simply use the one measured in the 
simulations, since it is guaranteed to satisfy~(\ref{pdfnorm}).  
Whenever we do so, we will also show the extent to which this 
is self-consistent by showing the ratio of the left hand side 
to the right hand side of equation~(\ref{meaneul}).  

\begin{figure}
\centering
\mbox{\psfig{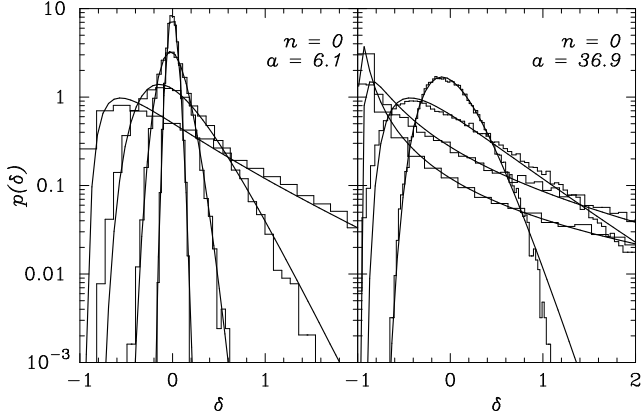}}
\caption{The Eulerian space probability distribution function 
$p(\delta)$ as a function of overdensity $\delta$, for clustering 
from white noise initial conditions.  
Each panel shows four choices of scale 
$R/L = 0.02$ (broadest curves), $0.4$, $0.8$ and $0.16$ 
(narrowest curves).  
Histograms show the distribution measured in the simulations; 
thicker, smoother curves show Generalized Inverse Gaussian 
distributions that have the same variance. }
\label{pdf0e}
\end{figure}

\begin{figure}
\centering
\mbox{\psfig{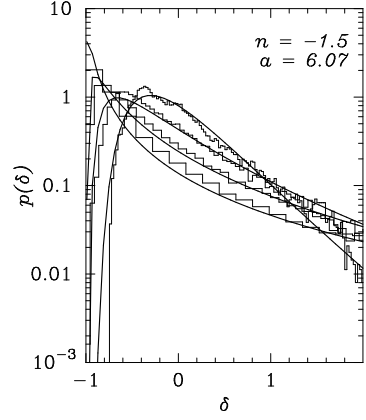}}
\caption{Same as the previous figure, but for clustering from 
$n=-1.5$ initial conditions.   }
\label{pdf1he}
\end{figure}

Figs.~\ref{pdf0e} and~\ref{pdf1he} show the Eulerian space 
probability distribution function for a range of cell sizes.  
The histograms show the $p(\delta)$ distribution measured in 
the simulations.  Solid curves show Generalized Inverse Gaussian 
distributions (e.g. Sheth 1998) that have the same variance:
\begin{equation}
p(\delta)\,{\rm d}\delta =
{s^{-\lambda}\over 2K_\lambda(\omega)}\,
{\rm e}^{-{\omega\over 2}(s+s^{-1})}\ {{\rm d}s\over s},\qquad
\label{ginvg}
\end{equation}
where $s = 1/(1+\delta)^{(n+3)/3}$, 
$K_\lambda(\omega)$ is a modified Bessel function of the 
third kind, and $\lambda = -3/[2(n+3)]$, if the initial power 
spectrum had slope $n$.  The parameter $\omega$ is related to 
the variance by the relation
\begin{equation}
\langle(1+\delta)^2\rangle \equiv 1+\bar\xi_{\rm m} = 
K_{3\lambda}(\omega)/K_\lambda(\omega)  
\end{equation}
(since $\langle\delta\rangle=0$, and $K_{\lambda}=K_{-\lambda}$).  
For the curves shown, the values of $\bar\xi_{\rm m}$ are as follows:  
when $n=0$ and $a=6.1$, 
then $\bar\xi_{\rm m}=0.62$, 0.1, 0.01, and 0.002 
for $R/L=0.02$, 0.04, 0.08 and 0.16, respectively.  
When $n=0$ and $a=37$, then the corresponding values of 
$\bar\xi_{\rm m}$ have grown to 10.9, 2.1, 0.4 and 0.07.  
When $n=-1.5$ and $a=6.07$, then $\bar\xi_{\rm m}=14$, 3.6, 0.98, 
and 0.26.  Thus, on small scales, the clustering is reasonably well 
evolved.  The figures show that the analytic formulae provide a 
reasonably good, but by no means perfect, fit to the simulation 
data on all scales.  The fit appears better on a log scale than 
on a linear scale.  Nevertheless, they will be used as convenient 
fitting functions to the Eulerian space distributions when 
they are used in Section~\ref{xihe}.  


\subsection{Biasing in Eulerian space}
This subsection compares the bias relation between haloes and 
mass measured in the simulations in the Eulerian space with the 
theoretical model developed in the previous sections.  
The theoretical model combines the Lagrangian expressions derived 
in Sections~\ref{wnics} and~\ref{scfics} with the Mo \& White (1996) 
model of Eulerian evolution discussed in Section~\ref{scmow}.  
However, it is independent of the Eulerian space dark matter 
distribution function.  

\begin{figure}
\centering
\mbox{\psfig{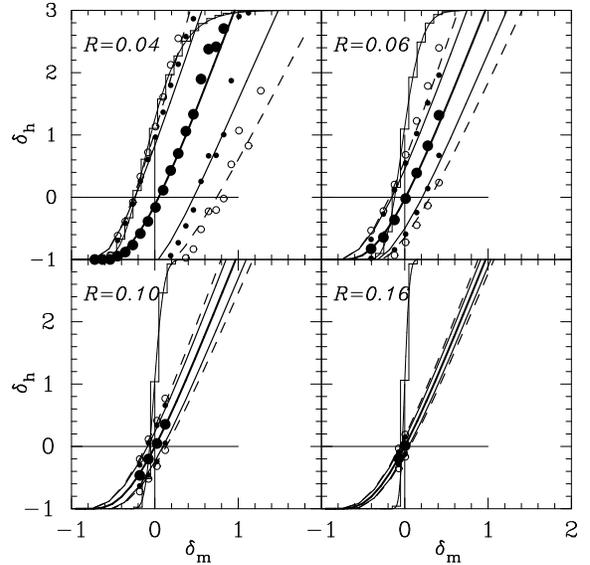}}
\caption{The Eulerian space bias relation for haloes containing 
more than $m=32$ particles that form from white noise initial 
conditions.  
Symbols show quantities measured in the simulations: 
large filled circles show the mean, 
smaller filled circles show the rms scatter, 
and open circles show the scatter if the halo counts were Poisson.  
Curves show the model predictions.  
Haloes were identified at an expansion factor of $a=6.1$; 
the bias relation was computed from the halo-centre-of-mass and 
mass distributions at that time.  The histograms that rise from 
left to right in each panel show the cumulative counts-in-cells 
distribution.  The simulations provide a good test of the theory 
only in the range where these curves are steep.  The solid lines 
through the histograms show the cumulative Generalized Inverse 
Gaussian distribution fitting functions.  }  
\label{bias0e3}
\end{figure}

\begin{figure}
\centering
\mbox{\psfig{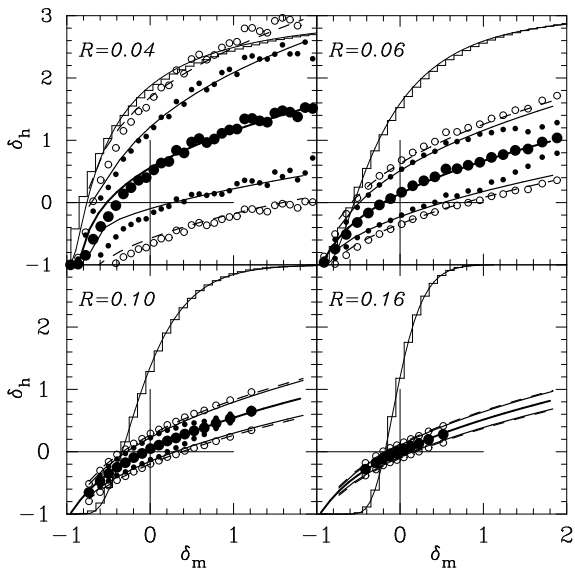}}
\caption{The same as the previous Figure, i.e., $n=0$ and $m=32$, 
but here $a=36.9$.  }
\label{bias0e7}
\end{figure}

\begin{figure}
\centering
\mbox{\psfig{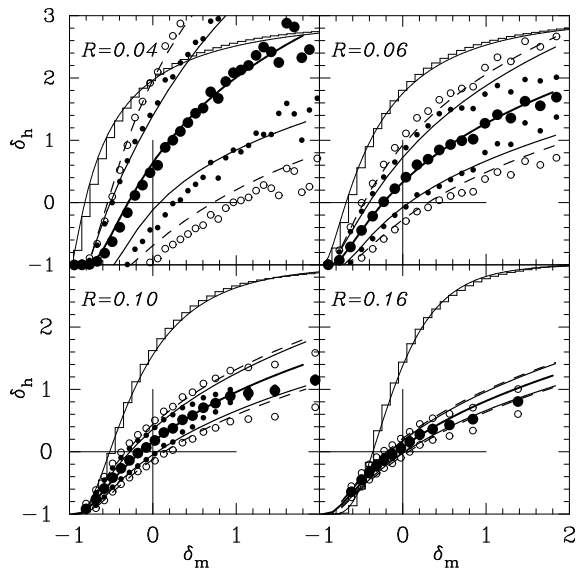}}
\caption{The same as the previous Figure, but for $n=-1.5$, $m=32$, 
 and $a=6.07$.  }
\label{bias1e7}
\end{figure}

Figs.~\ref{bias0e3}---\ref{bias1e7} show the bias relation 
for the same haloes as in previous figures, but now the 
mean and the scatter are measured in Eulerian space.  
The histograms show the cumulative Eulerian space distribution 
function, and the solid lines through the histograms show the 
cumulative Generalized Inverse Gaussians that have the same variance.  
As in the Lagrangian case, these cumulative curves are included to 
show the range over which the simulations provide a good test of 
the theory; this range is where the cumulative curves are steep.  
The figures show that the theoretical curves for the mean 
Eulerian bias fit the corresponding quantities measured in the 
simulations very well.  This agreement has already been shown 
by Mo \& White (1996).  What is new here is that our expressions 
for the scatter around the mean bias relation appear to describe 
that measured in the simulations very well also.  The agreement at 
small $R$ is particularly gratifying, since there the scatter is 
significantly less than Poisson.  This shows that our model is 
able to account correctly for volume exclusion effects.  
The agreement between theory and simulation when $n=-1.5$ is 
also encouraging.  It suggests that our simple analytic model for 
quantifying the effects of volume exclusion is reasonably 
accurate even when the initial conditions are significantly 
different from white noise.  


\subsection{Dependence of the mass function on local overdensity}
\begin{figure}
\centering
\mbox{\psfig{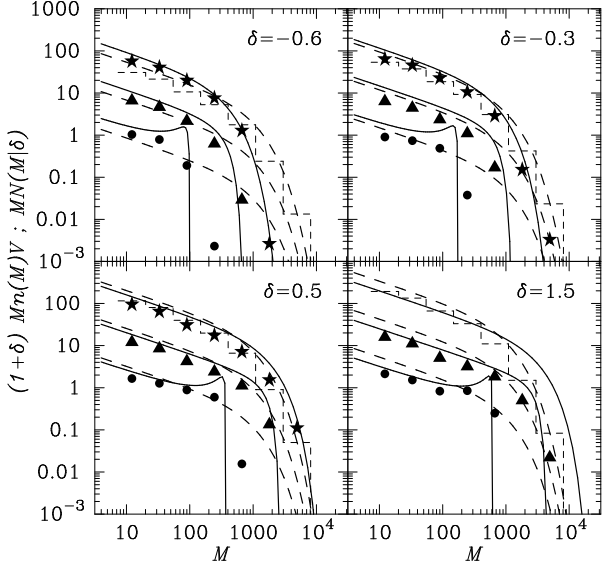}}
\caption{The mass function of haloes that are in Eulerian 
cells $V$ which have overdensity $\delta$.  
Haloes were identified at an expansion factor of $a=36.9$ 
in the simulations with white noise initial conditions.  
The three sets of curves in each panel show results for three cell 
sizes:  $R/L=0.04$ (bottom), $0.08$ and $0.16$ (top).  
Filled symbols show the average number of haloes in those 
Eulerian cells that have overdensity $\delta$.  
The histogram shows $(1+\delta)$ times the universal mass 
function times the largest cell size.  On a log-log plot, it has the 
same shape but a different amplitude for the other cell sizes.  
The dashed curves show the corresponding theoretical 
curves: $(1+\delta)V$ times the universal mass function.  
The solid curves show the mass function computed using the 
Mo \& White bias model of equation~(\ref{equalns}).}  
\label{lk0}
\end{figure}

\begin{figure}
\centering
\mbox{\psfig{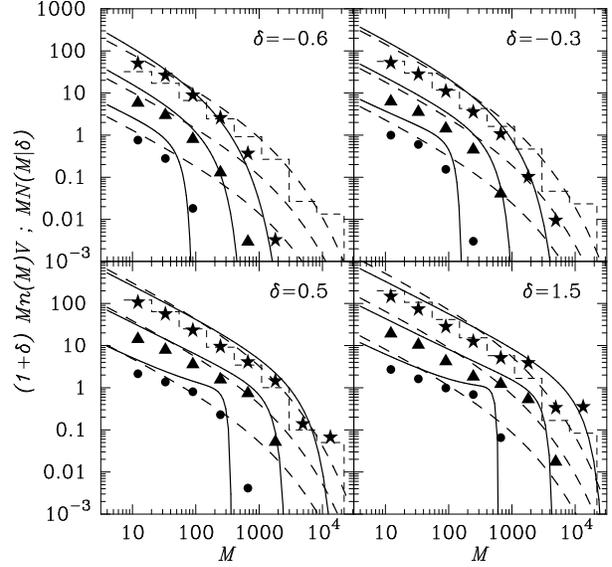}}
\caption{Same as the previous figure, but for $n=-1.5$ initial 
conditions, and an expansion factor of $6.1$.  }
\label{lk1h}
\end{figure}

There is another way to show that the Mo \& White Eulerian space 
bias model is reasonably accurate.  Equation~(\ref{meaneul}) shows 
that the unconditional, universal mass function $n(M)$ is simply 
related to the conditional mass function $N(M|\delta)$ of haloes that 
are known to be in Eulerian cells $V$ which have overdensity $\delta$ 
averaged over all values of $\delta$.  
In the Mo \& White model, $N(M|\delta)$ is given by 
equation~(\ref{equalns}); in general, it is different from 
$(1+\delta)\,n(M)V$.  In particular, in the model, the shape of the 
mass function depends on the Eulerian overdensity:  the ratio of 
massive haloes to less massive haloes is larger in dense regions 
than in less dense regions.  Figs.~\ref{lk0} and~\ref{lk1h} show that 
this is consistent with what is measured in the simulations.  

These figures are similar to Fig.~1 of Lemson \& Kauffmann (1999).
They show the conditional mass function $N(M|\delta)$ 
for haloes in Eulerian cells $V$ that have overdensity $\delta$, 
for a range of choices of $\delta$ and $V$.  The top left panel 
shows the range $-0.8\le\delta\le -0.4$, the top right shows 
$-0.5\le\delta\le -0.1$, bottom left shows $0.3\le\delta\le 0.7$ 
and bottom right shows $1.2\le\delta\le 1.8$.  The three 
sets of curves in each panel show different cell sizes:  
$R/L = 0.04$ (bottom). $0.08$ and $0.16$ (top).  
The histograms show $(1+\delta)$ times the largest cell 
volume $V$ times the unconditional mass function measured in the 
simulations.  The associated dashed curves show $(1+\delta)V$ times 
the Press--Schechter formula for the universal unconditional mass 
function with $\delta_{\rm c}=1.7$.  The dashed curves provide 
good but not perfect fits to the histograms.  Changing the cell 
size on a log-log plot simply changes the amplitude of the 
curves, so for smaller cell sizes we only show the analytical 
formula.  The solid symbols show the actual conditional mass 
function measured in the simulations and the bold curves show 
the conditional mass function of equation~(\ref{equalns}).  
The symbols differ from the histograms in the same way that the 
solid curves differ from the dashed curves.  
(The bottom right panel has only two sets of symbols because there 
were no large cells with the given range in $\delta$.) 
This shows explicitly that, just as the Press--Schechter formula 
provides a reasonable fit to the unconditional mass function 
averaged over all Eulerian cells, the Mo \& White model provides 
a reasonable fit to the mass function if only cells of a 
certain density range are used when computing the average.  

The data points show the mean number of haloes in cells $V$ 
that are known to have overdensity $\delta$.  Since not all 
cells have the same number of haloes, there is some scatter 
around this mean.  Our extension of the Mo \& White model allows 
us to predict the rms `error bars' on the data points.  
We have not shown them here.  

\subsection{Eulerian space halo correlation functions}\label{xihe}
This subsection compares the Eulerian space halo--mass and halo--halo 
correlations measured in the simulations with the theoretical model 
developed in the previous sections.  
To do this requires knowledge of the distribution function of 
the probability that a randomly placed Eulerian cell of size $V$ 
contains mass $M$.  Although Section~\ref{scmow} discussed how the 
Mo \& White approach can be extended to derive this distribution 
self-consistently, here we simply follow the approach used by 
Mo \& White.  Namely, we will use the Eulerian probability 
distribution functions measured in the simulations themselves 
(and, in fact, we will use the Generalized Inverse Gaussian fits 
to these distributions), rather than the ones required by 
self-consistency.  
Recall that this means that there is no longer any guarantee that 
the model gives the correct number density of haloes.  Below, 
we will show explicitly that the model is not self-consistent 
on small scales.  

\begin{figure}
\centering
\mbox{\psfig{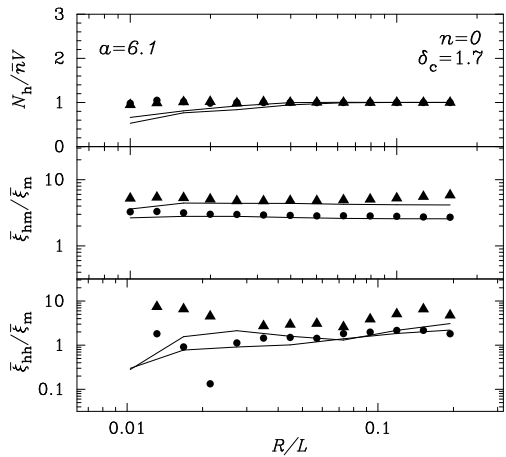}}
\caption{Various Eulerian space quantities as a function of 
Eulerian cell size.  Top panel shows $N(>\!m|V)/n(>\!m)V$, middle 
panel shows $\bar\xi_{\rm hm}(>\!m|V)/\bar\xi_{\rm m}$, and 
bottom panel shows $\bar\xi_{\rm hh}(>\!m|V)/\bar\xi_{\rm m}$.  
Filled circles, triangles, squares and stars show results for 
haloes in the simulations that contain more than 32, 64, 128, 
and 256 particles, respectively.  Solid curves show the model 
predictions.  }
\label{xi0e3}
\end{figure}

\begin{figure}
\centering
\mbox{\psfig{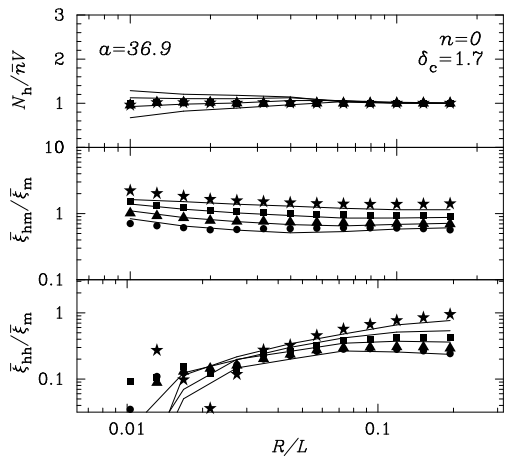}}
\caption{Same as the previous figure, but for haloes identified 
at a later output time.  }
\label{xi0e7}
\end{figure}

\begin{figure}
\centering
\mbox{\psfig{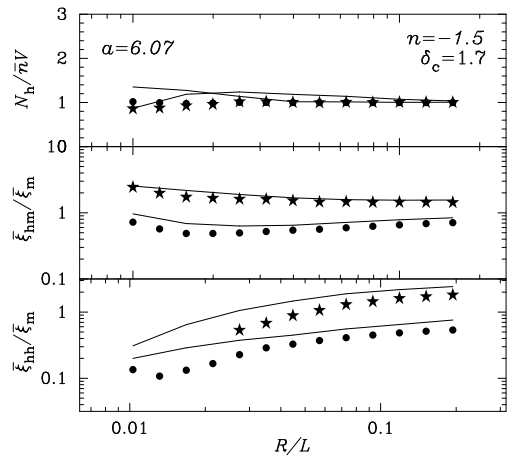}}
\caption{Same as the previous figure, but for initial conditions 
with a power spectrum with slope $n=-1.5$.  }
\label{xi1he7}
\end{figure}

Figs.~\ref{xi0e3}--\ref{xi1he7} show the result of comparing 
the Mo \& White model with the Eulerian space distributions 
measured in the simulations.  The top panels in each figure 
show $N(>\!m|V)/n(>\!m)V$, the middle panels show 
$\bar\xi_{\rm hm}(>\!m|V)/\bar\xi_{\rm m}(V)$, and the bottom 
panels show $\bar\xi_{\rm hh}(>\!m|V)/\bar\xi_{\rm m}(V)$ 
as a function of Eulerian scale.  The symbols show the quantities 
measured in the simulations, and are coded similarly to those 
in the corresponding Lagrangian space plots.  The solid curves 
show the theoretical quantities.  

If the Mo \& White model were self-consistent, then the theoretical 
curves in the top panels of each figure would be unity on all 
scales.  Thus, the figures show that the Mo \& White model is 
inconsistent on small scales.  The middle panels show that, 
despite this inconsistency, the model provides a good fit to 
the Eulerian space cross correlation between haloes and mass.  
This is primarily a consequence of the fact that the mean Eulerian 
bias is well reproduced by the Mo \& White model 
(Figs.~\ref{bias0e3}---\ref{bias1e7}).  These curves are similar 
to those shown in Fig.~4 of Mo \& White (1996).  
The bottom panels should be compared with Fig.~5 of 
Mo \& White (1996).  Whereas their model curves increase as 
$R/L$ decreases, ours do not.  Thus, our model for the volume 
averaged halo--halo correlation function works significantly better 
than the one they used.  This is to be expected, since our model 
explicitly takes account of volume exclusion effects, whereas 
theirs did not.  
The bottom panels also show that, on sufficiently large scales, 
one consequence of dynamical evolution is to make massive haloes 
more strongly clustered than less massive ones.  
This is in agreement with earlier predictions 
(Cole \& Kaiser 1989; Mo \& White 1996) as well as with the 
model developed here.  

\section{Discussion}
Numerical simulations show that haloes are biased tracers of the 
matter distribution.  This bias depends nonlinearly on scale and 
on halo mass, and the bias on any given scale is stochastic.  
This paper describes an analytic model which describes this 
nonlinear, stochastic biasing, as well as its evolution, 
reasonably accurately.  

The model is consistent with the assumption that disconnected 
volumes in the initial Lagrangian space may be treated as being 
mutually independent.  This assumption allows one to use quantities 
associated with the merger histories of dark haloes to estimate 
the Lagrangian space correlation functions of these haloes.  
The assumption of indepedence is most likely to be accurate if 
the initial distribution was Poisson or Gaussian white noise.  
The Poisson model is described in detail in Appendix~\ref{pics}, 
where various subtle issues involved in this approach are 
discussed rigourously.  In the limit of small fluctuations and 
large numbers of particles, statements about clustering from 
Poisson initial conditions are easily related to those that 
describe clustering from white noise initial conditions 
(Sheth 1995, 1996).  


Section~\ref{wnics} showed these expressions for the mean and higher 
order moments of the halo distribution, for white noise initial 
conditions.  The final expressions compliment and extend those 
derived by Mo \& White (1996).  In particular, the results of this 
section allow one to account for volume exclusion effects which 
arise from the fact that haloes initially occupy a volume that is 
proportional to their mass.  These effects were described, but not 
quantified by Mo \& White.  Our results also include the effects 
of the scatter among different formation histories of individual 
regions in the initial conditions on the statistics of the halo 
distribution in space---another effect that was described, 
but not quantified, by Mo \& White.  

Whereas disconnected volumes are mutually independent in the white 
noise case, this is not true for more general Gaussian initial 
conditions.  However, Sheth \& Lemson (1998) showed that it is 
possible to provide a good approximate description of the forest of 
merger history trees associated with haloes which form from initial
conditions with large scale correlations by simply ignoring these 
correlations.  In the Mo \& White model, knowledge of the merger 
history trees is equivalent to knowledge of the spatial distribution 
of dark haloes.  Section~\ref{scfics} used this fact to argue that 
the white noise results could be used to provide simple analytic 
approximations for the higher order moments of the Lagrangian space 
halo distribution even when the initial power on large scales is 
substantial.  The Sheth \& Lemson merger tree results suggest that 
these analytic approximations should also be reasonably accurate.  

As a result of dynamical evolution, the evolved halo distribution 
is different from that in the initial Lagrangian space.  To describe 
the evolved distribution we used the spherical model, in the way 
suggested by Mo \& White, to relate the initial halo distribution 
described above to the final evolved one.  
We showed that in addition to allowing one to estimate the 
evolved halo--mass and halo--halo correlation functions, the 
Mo \& White model could have been used to compute the Eulerian 
space probability distribution function of the dark matter itself.  
This is a potentially useful extension of their model.  

Once the model had been fully specified, we compared it with 
numerical simulations of hierarchical gravitational clustering.  
Comparison with the halo distribution in the simulations 
(Section~\ref{sims}) showed that while the Mo \& White bias model 
is reasonably accurate when describing the mean Lagrangian space 
bias relation of massive haloes, it predicts the wrong mean value 
for less massive objects.  
Our extension of the Mo--White model allows us to compute the 
higher order moments of the bias relation.  For massive objects 
(those for which the Mo--White mean is accurate), it describes the 
scatter around the mean well.  For less massive objects, when 
the Mo--White model gets the mean value wrong, our model for 
the scatter around the mean is still in qualitative agreement 
with the simulations.  

Results for the halo distribution in Eulerian space were more 
encouraging.  The Mo \& White model describes the mean properties of 
the bias relation in Eulerian space well, for a larger range of 
masses than in the Lagrangian space, and our extension of the model 
is able to describe the scatter around this mean well.  
Our model works even on scales where volume exclusion effects are 
important.  This is very encouraging, since our model provides 
simple, analytic expressions for these higher order moments.  
Although our simulations do not have the dynamic range to 
investigate a large mass range, those of Jing (1998) do.  
Jing finds that on large scales where 
$\bar\xi_{\rm hh}^{\rm E}/\bar\xi_{\rm m}$ is constant, 
the Eulerian space low--mass halo distribution is more clustered 
than the Mo \& White model predicts.  In other words, he finds that 
the large scale mean bias relation between low--mass haloes and 
the mass is larger than the mean bias relation that the Mo \& White 
model predicts.  It is interesting that this is the same trend 
we found in our study of the Lagrangian space halo distribution.  
This has an important consequence.  

The Mo \& White model has two parts:  
the first is a model of the initial number density 
and spatial distribution of haloes, 
and the second models their subsequent dynamical evolution.  
Given only Jing's result, one might have thought that the Mo \& White 
fails only in the second step; that using the spherical model to 
translate from Lagrangian to Eulerian space is inaccurate.  
If so, one might have thought that the Zel'dovich approximation, 
or variants of it, could be combined with the initial distribution 
described here to derive accurate estimates of the evolution of the 
spatial distribution of massive as well as less massive haloes.  This 
is the sort of approach taken by Catelan, Mataresse \& Porciani (1998).  
To date, they have only studied the halo distribution on scales 
larger than that of a typical halo, since their approach does not 
allow them to account for the effects of volume exclusion.  
Since we are able to account for volume exclusion, it may be 
interesting to combine some of the results presented here with 
their work.  

However, our results show that the Mo \& White model fails in 
Lagrangian space:  it does not describe the initial spatial 
distribution of low mass haloes correctly.  This is not so 
surprising, since it is well known that the spherical model for the 
collapse of haloes, on which the first step of the Mo \& White model 
is based, is more likely to be accurate for massive objects than for 
less massive ones (e.g. Bernardeau 1994).  
If it is not so much the spherical model of the evolution of the 
halo distribution, but rather the spherical collapse model 
for the formation of small mass haloes itself that is wrong, 
then we expect the discrepancy Jing measures for the Eulerian 
space distribution of the haloes in his simulations to be reflected 
in the shape of the unconditional mass function.  The mass function 
in the simulations does indeed differ from the Press--Schechter 
function, and this difference is in the correct sense:  whereas the 
theory predicts approximately the correct number of massive haloes, 
there are fewer low mass haloes in the simulations than the 
Press--Schechter formula predicts.  Quantifying this relation between 
the unconditional mass function and the large scale bias relation is 
the subject of ongoing work.  

In this paper we have gone to a fair amount of trouble to derive 
a realistic, accurate, analytic model for the scatter in the 
halo-to-mass bias relation.  This is because knowledge of this 
scatter allows one to address a number of interesting problems, 
some of which we list briefly below.  
To relate these results to the observed distribution of galaxies 
is complicated.  Galaxies are thought to form inside dark 
matter haloes (White \& Rees 1978; White \& Frenk 1991).  
Semi-analytic models of this galaxy formation process 
(e.g. Kauffmann, White \& Guiderdoni 1993) show that the number 
of galaxies which form in a given dark matter halo is stochastic.  
Lemson \& Kauffmann (1999) showed that most of the physical 
parameters of a dark matter halo on which galaxy formation processes 
are expected to depend, while they may depend on the halo mass, 
are independent of the halo's environment.  
Thus, their results suggest that quantities like the average number, 
or the scatter in this number, of galaxies in a dark matter halo 
ultimately depend on the halo mass.  
So it should be possible to provide semi-analytic estimates of 
the mean galaxy-number-to-halo-mass bias relation, as well as the 
scatter in this relation.  
When combined with our results for the mean and higher order 
moments of the bias between dark matter haloes and the underlying 
matter distribution, such a relation would allow one to relate the 
observed galaxy distribution to that of the underlying dark matter 
distribution.  Thus, our expressions for the scatter in the 
halo--dark matter bias relation can be used to extend the results 
of Kauffmann, Nusser \& Steinmetz (1997) to smaller scales.  
In addition, combining the galaxy number to halo mass bias 
relation with the dark halo to dark matter bias relation may allow 
one to compute estimates of the expected scatter in the Tully-Fisher 
relation, to study the bias associated with estimating $\Omega_0$ 
from redshift distortions (Pen 1998; Dekel \& Lahav 1998), to evaluate the 
compatibility between observations of the number density and 
correlation functions of objects at high redshift and various 
cosmological models (Mo, Mao \& White 1998), and to model 
the evolution of the cluster--cluster correlation function in 
different cosmological models (Mo, Jing \& White 1997).  

This paper has dealt primarily with the problem of quantifying 
the mean and higher order moments of the halo bias given the 
matter fluctuation field 
(e.g. $\langle \delta_{\rm h}|\delta_{\rm m}\rangle$).  
The inverse problem is equally, if not more, interesting.  
The problem of estimating the mean and higher order moments of 
the matter fluctuation field given the halo distribution 
(e.g. $\langle \delta_{\rm m}|\delta_{\rm h}\rangle$).  
is the subject of ongoing work.  

\section*{Acknowledgments}
We thank H. Mo \& S. White for many helpful conversations, and 
for providing access to their numerical simulations, and  
A. Heavens for a prompt and helpful referee report.  
The analytic results presented here were derived during the 
spring of 1996.  RKS thanks Avishai Dekel and the Racah Institute 
of Physics at the Hebrew University, Jerusalem for their warm 
hospitality and financial support during that time.  
GL thanks T. Banday for revealing two issues, while visiting 
Luxor, which provided the inspiration to complete this work.

\appendix
\section{Poisson initial conditions}\label{pics}
In this Appendix the excursion set approach is used to derive 
expressions for the unconstrained mass function, and then  
for the constrained mass function and associated merger 
probabilities.  The approach follows and 
extends that of Epstein (1983) and Sheth (1995) in the following 
way.  These earlier analyses considered spherical collapse around 
particles in the initial Poisson distribution.  However, in this 
paper we want to compute averages over all randomly placed cells 
in the initial distribution, not just those that are centred on 
particles.  So, in the next few subsections, we derive expressions 
for the constrained and unconstrained mass functions where the 
restriction to volumes centred on particles has been dropped.  
It turns out that the modification to the previously derived 
expressions is trivial.  Therefore, the first two subsections may 
seem a little pedagogical---we have included them to set notation.  
Readers familiar with the Poisson excursion set analysis may 
prefer to skip directly to Section~\ref{fnhalo}.  

The spatial distribution of these haloes, in the initial Lagrangian 
space, is described in Sections~\ref{fnhalo}--\ref{hom}.  
Comparison of these results with numerical simulations is often 
done for haloes having a range of masses.  There is some subtlety 
in doing this correctly---this is discussed in Section~(\ref{range}).  
That all these Poisson results are easily extended to describe 
clustering from white noise initial conditions is shown in 
Section~\ref{wnlim}.  Essentially, those statements about clustering 
from white noise initial conditions which are known (e.g., the 
conditional and unconditional mass functions, and the mean 
bias relation), can be derived by taking appropriate limits of 
the corresponding Poisson statements.  The same limiting procedure 
can be used to derive statements about the higher order moments of 
the Lagrangian space halo distribution.  It is these expressions 
that are presented in the main text.  

\subsection{The unconstrained mass function}\label{umf}
Consider a Poisson distribution of particles with mean density 
$\bar n$.  This means that a volume of size $V$ placed at a random 
position in this distribution will contain exactly $N$ particles 
with probability 
\begin{equation}
p(N,V) = {(\bar nV)^N\,{\rm e}^{-\bar nV}\over N!}.
\label{pois}
\end{equation}
Furthermore, if it is known that there are $N$ particles in $V_2$, 
then the probability that there are $j$ particles in $V_1$ placed 
randomly within $V_2$ is 
\begin{eqnarray}
p(j,V_1|N,V_2) &=& {p(j,V_1)\ p(N-j,V_2-V_1)\over p(N,V_2)} \nonumber \\
&=& {N\choose j}\, \left({V_1\over V_2}\right)^j\,
\left(1 - {V_1\over V_2}\right)^{N-j}.
\label{pjoint}
\end{eqnarray}

Now choose a random position in the distribution, and compute the density 
within concentric spheres centred on this position.  Call the curve traced 
out by the number of particles contained within a sphere $V$ centred 
on this point, as a function of the sphere size $V$, a trajectory.  
Then each position in the Poisson distribution has its associated 
trajectory.  Let $f^{\rm e}(\delta_1)$ denote the probability that, 
for all concentric spheres centred on a randomly chosen position, the 
density never exceeds the threshold value $\bar n(1+\delta_1)$.  One 
way to compute this probability is to compute the fraction of 
trajectories for which $N(V) < \bar nV(1+\delta_1)$ for all $V$, 
where $N(V)$ is the number of particles within $V$.  This quantity 
can be computed as follows.  

Start with an arbitrarily small sphere centred at the chosen position, and 
consider successively larger concentric spheres.  As the volume increases by 
an infinitesimal amount, the number of particles contained within the current 
sphere either remains the same, or increases by one.  (Strictly speaking, the 
probability that the number of particles increases by one is an infinitesimal, 
the probability that the number increases by two is an infinitesimal of higher 
order, an increase by three particles is an infinitesimal of still higher 
order, and so on.)  Therefore, a given value of $\delta_1$ defines a series 
of volumes $V_1< V_2< \ldots$ for which  
\begin{equation}
j/V_j = \bar n(1+\delta_1) \equiv \bar n/b_1.
\label{vj}
\end{equation}
The final equality defines $b_1 = 1/(1+\delta_1)$, a parameter which will 
be useful later.  The quantity of interest, $f^{\rm e}(\delta_1)$, is one 
minus the probability that $V_j$ is the largest sphere centred at the 
chosen position that has density $\bar n(1+\delta_1)$, summed over 
all $V_j$.  That is, 
\begin{equation}
1 - f^{\rm e}(\delta_1) = 
\sum_{j=1}^\infty p(j,V_j)\,f^{\rm e}(\delta_1|j,V_j) ,
\end{equation}
where the first term in the sum is the probability that $V_j$ contains 
exactly $j$ particles, and the second term expresses the probability 
that, given that $V_j$ contains exactly $j$ particles, no concentric sphere 
larger than $V_j$ is denser than it.  Epstein (1983) shows that 
\begin{equation}
f^{\rm e}(\delta_1|j,V_j) = {\delta_1\over 1+\delta_1} = 1-b_1 ,
\label{fext}
\end{equation}
and he discusses why it is independent of $j$.  Thus, 
\begin{equation}
1 - f^{\rm e}(\delta_1) = (1-b_1)\sum_{j=1}^\infty p(j,V_j) = b_1,
\end{equation}
where the sum is simplified by recognizing that it is $b_1$ times the first 
moment of the Borel distribution (Borel 1942).  This shows that 
$f^{\rm e}(\delta_1) = 1-b_1$, so that it is the same as 
$f^{\rm e}(\delta_1|j,V_j)$.  This is simply a consequence of the fact 
that, since the distribution is Poisson, the probability that all 
larger volumes containing a given volume are less dense than a given 
value depends only on the density within the volume, and not on the 
number or the distribution of the particles within it.  

The expression above implies that the probability that 
at least one sphere centred on a randomly chosen position in a Poisson 
distribution is denser than $\bar n(1+\delta_1)$ is $b_1$.  In other words, 
of the infinity of spatial positions in a Poisson distribution, and, of 
the infinity of associated trajectories, only a fraction $b_1$ are at the 
center of at least one sphere that is denser than $\bar n(1+\delta_1)$.  
That is, only a fraction $b_1$ of the trajectories ever have 
$N(V) \ge \bar nV(1+\delta_1)$ for at least one value of $V$.  

Let $F(j,b_1)$ denote the fraction of trajectories for which 
$N(V_j) = j$, and for which $N(V_k)<k$ for all $V_k>V_j$.  Then 
\begin{equation}
F(j,b_1) = p(j,V_j)\,f^{\rm e}(\delta_1|j,V_j) ,
\label{fjb}
\end{equation}
where the first term gives the probability that a trajectory has 
$N(V_j) = j$, and the second term gives the probability that $V_j$ 
is the largest volume at which the trajectory exceeds the threshold
$\bar n(1+\delta_1)$.  

There is a useful relation between equations~(\ref{pois}), 
(\ref{pjoint}) and~(\ref{fjb}).  Let $_2\!V_k \equiv kb_2/\bar n$.  
Then 
\begin{equation}
p(j,V) = \sum_{k=j}^\infty p(j,V|k,\,_2\!V_k)\,F(k,b_2) ,
\label{pjfk}
\end{equation}
provided $j/V\ge \bar n$, and $b_2\le b_1\equiv \bar nV/j$.  
To see this, note that the left hand side includes all trajectories 
that have value $j$ at $V$.  Suppose each trajectory is labelled by 
the value of $k$ for which $_2\!V_k$ is the largest volume at which 
that trajectory crossed the line $\bar n(1+\delta_2)$.  
Trajectories which cross the line for the final time with value 
less than $j$ cannot also pass through $V$ with value $j$.  Therefore, 
the sum on the right hand side is only over those trajectories that 
cross the line $\bar n(1+\delta_2)$ for the final time with $k\ge j$, 
and also pass through $V$ with value $j$.  Clearly, the left hand 
side must equal the right.  Direct substitution shows that 
equations~(\ref{pois}), (\ref{pjoint}) and~(\ref{fjb}) do satisfy 
this relation.  The normalization and first moment of Consul's (1989) 
generalized Poisson distribution 
aid in proving this result.  

Define an isolated region as a spherical region within which the average 
density is $\bar n (1+\delta_1)$, and for which the average density 
within all larger concentric spheres is less than this.  Then 
$F(j,b_1)$ denotes the fraction of space that is associated with 
isolated $(j,b_1)$-volumes.  If $N(j,b_1)$ denotes the number of such 
volumes, and $V_{\rm U}$ denotes the total volume, then 
\begin{displaymath}
F(j,b_1) = N(j,b_1)\,{V_j\over V_{\rm U}} = 
{N(j,b_1)\over V_{\rm U}}\,{jb_1\over \bar n},
\end{displaymath}
so that the number density $\bar n(j,b_1)$ of such isolated 
volumes is 
\begin{eqnarray}
\bar n(j,b_1) &\equiv& {\bar n\over jb_1}\,F(j,b_1) = 
\bar n(1-b_1)\,{(jb_1)^{j-1}{\rm e}^{-jb_1}\over j!} \nonumber \\
&=& \bar n(1-b_1)\,\eta(j,b_1),
\label{borel}
\end{eqnarray}
where $\eta(j,b_1)$ is the Borel$(b_1)$ distribution.  
Thus, the Borel$(b_1)$ distribution gives the probability that an 
isolated region contains exactly $j$ particles [since 
$\sum_j \eta(j,b_1) = 1$, and $\sum_j j\,\eta(j,b_1) = 1/(1-b_1)$].  

Following Bond et al. (1991), it will be convenient to associate 
these isolated regions with collapsed haloes.  Then 
equation~(\ref{borel}) is the unconditional mass function, since 
it gives the number density of collapsed objects that contain 
exactly $j$ particles.  

It is interesting to compare equations~(\ref{fjb}) and~(\ref{borel}) 
with the results of Epstein (1983).  In his analysis, Epstein only 
considered those trajectories that were certainly centred on 
particles of the Poisson distribution.  Here, that restriction has 
been dropped.  Let $f(j,b_1)$ denote the fraction of trajectories 
that are centred on particles and are associated with isolated 
regions containing exactly $j$ particles.  Epstein's expression 
for $f(j,b_1)$ implies that 
\begin{equation}
F(j,b_1) = b_1\,f(j,b_1) .
\label{effs}
\end{equation}  
Thus, the effect of considering the set of all trajectories, rather than 
the subset that are centred on particles, is simply to introduce 
the $b_1$ term.  This is sensible.  In the limit in which the 
threshold $\delta_1\to\infty$, $b_1\to 0$.  In this limit, the only 
trajectories that ever exceed the threshold are those that are 
centred on particles, and they exceed the threshold only when the 
volume is vanishingly small.  In this limit, $f(j,b_1)=1$ if $j=1$, 
and it is zero otherwise.  On the other hand, the subset of 
trajectories that are centred on particles is a vanishingly small 
fraction of the set of all trajectories, so that, as 
$\delta_1\to\infty$, the fraction of all trajectories that ever 
exceed $\delta_1$ tends to zero.  So, in this limit, 
$F(j,b_1)\to 0$ for all $j$.  

\subsection{The constrained mass function}\label{cmf}
The probability that a randomly placed volume $_1\!V_j$ contains 
exactly $j$ particles and has density $\bar n(1+\delta_1)$, and 
that the larger volume $_2\!V_k >\,_1\!V_j$ including $_1\!V_j$ 
contains exactly $k$ particles, has density $\bar n(1+\delta_2)$, 
and is isolated, is 
\begin{displaymath}
p(j,\,_1\!V_j)\ p(k-j,\,_2\!V_k -\,_1\!V_j)\ 
f^{\rm e}(\delta_2|k, _2\!V_k).
\end{displaymath}
Equation~(\ref{fext}) shows that 
$f^{\rm e}(\delta_2|k, _2\!V_k) = (1-b_2)$.  
The probability $F(j,b_1|k,b_2)$ that $_1\!V_j$ is itself isolated within 
the isolated region $_2\!V_k$ [that is, the average density within all 
volumes $V$ that include $_1\!V_j$ and are within $_2\!V_k$ is less 
than $\bar n(1+\delta_1)$] satisfies a recursion relation:
\begin{eqnarray}
F(j,b_1|k,b_2) \!\!\!\!&=& \!\!\!\!
{p(j,\,_1\!V_j,k,\,_2\!V_k)\,f^{\rm e}(\delta_2|k, _2\!V_k)\over
F(k,b_2)} 
\nonumber \\ \!\!\!\!& &\!\!\!\!
 - \sum_{m>j}^k F(m,b_1|k,b_2)\,p(j,\,_1\!V_j|m,\,_1\!V_m).
\label{recur}
\end{eqnarray}
The numerator in the first term on the right is the joint probability 
above, in which $_1\!V_j$ is not necessarily isolated.  The denominator 
is included since it is known that $_2\!V_k$ is isolated.  
From this first term, we must subtract the probability that a volume 
$_1\!V_m$ containing $_1\!V_j$ was itself the largest isolated region within 
$_2\!V_k$.  This is just the product of the probability $F(m,b_1|k,b_2)$ 
times the probability $p(j,\,_1\!V_j|m,\,_1\!V_m)$ that there were exactly 
$j$ particles within $_1\!V_j$ given that they were within the isolated 
region $_1\!V_m$, summed over all $m$ larger than $j$.  Now, 
\begin{eqnarray}
p(j,\,_1\!V_j|m,\,_1\!V_m) &\equiv& 
{p(j,\,_1\!V_j)\ p(m-j,\,_1\!V_m-\,_1\!V_j)\over p(m,\,_1\!V_m)}
\nonumber \\
&=& {m\choose j}\left({_1\!V_j\over _1\!V_m}\right)^{j}
\left(1 - {_1\!V_j\over _1\!V_m}\right)^{m-j} \nonumber \\
&=& {m-1\choose j-1}
\left({j\over m}\right)^{j-1}\left(1 - {j\over m}\right)^{m-j} 
\label{pjbinom}
\end{eqnarray}
since $_1\!V_k = kb_1/\bar n$.  
This binomial-like term is necessary because not all configurations of 
particles that contribute to $F(m,b_1|k,b_2)$ will have had exactly $j$ 
particles within $_1\!V_j$.  Appendix~\ref{solvrec} shows that 
\begin{eqnarray}
F(j,b_1|k,b_2) &=& k\left(1-{b_1\over b_2}\right)\,
{k\choose j}\,{j^j\over k^k}\nonumber\\ 
& &\qquad\times\quad
\left(b_1\over b_2\right)^j \left(k-j{b_1\over b_2}\right)^{k-j-1}
\label{fjk}
\end{eqnarray}
satisfies the recursion relation given above.  

Let $f(j,b_1|k,b_2)$ denote the corresponding expression for 
volumes $_1\!V_j$ that are known to certainly be centred on a 
particle.  Then $f(j,b_1|k,b_2)$ is given by equation~(40) of 
Sheth (1995), and 
\begin{equation}
F(j,b_1|k,b_2) = (b_1/b_2)\,f(j,b_1|k,b_2). 
\label{efjs}
\end{equation}
Thus, as with the statements $F(j,b_1)$, the expressions for 
randomly placed volumes are easily related to those for volumes 
that are centred on particles.  The $(b_1/b_2)$ factor here 
plays the same role as the factor $b_1$ in equation~(\ref{effs}).  
It simply reflects the fact that, for a Poisson 
distribution, the particles within $_2\!V_k$ are distributed as 
though they are part of a Poisson distribution with average 
density $\bar n(1+\delta_2)$, rather than $\bar n$.  Moreover, 
the discussion in the final paragraph of section~\ref{umf} 
applies to the limiting behaviour of $F(j,b_1|k b_2)$ as 
$\delta_1\to\infty$, i.e., as $b_1\to 0$, just as it did for 
the limiting behaviour of $F(j,b_1)$.  

The similarity between $F(j)$ and $F(j|k)$ can be made still 
more striking.  Suppose there are $k$ particles in the volume 
$_2\!V_k$ and $j\le k$ particles in the subvolume $_1\!V_j$ within 
it.  Then $\bar n(1+\delta') = \bar n\,(k-j)/(kb_2-jb_1)$ is 
the density in the remaining volume $_2\!V_k - {_1\!V_j}$, and 
\begin{equation}
F(j,b_1|k,b_2) = 
\left({\delta_1-\delta'\over 1+\delta_1}\right)\ 
p(j, {_1\!V_j}|k, {_2\!V_k}) ,
\end{equation}
where $p(j|k)$ is given by equation~(\ref{pjoint}).  
Equation~(\ref{fjb}) shows that 
$F(k,b_1)$ is given by an analogous expression; there, 
the remaining volume is infinite, so that the overdensity 
in it, $\delta'$, is $0$ by definition.  Thus, $F(j|k)$ 
is related to $p(j|k)$ in the same way that $F(j)$ is related 
to $p(j)$.  


Recall that, although $F(j,b_1)$ differed from $f(j,b_1)$, the 
final expression for the number density of isolated $(j,b_1)$ volumes 
was the same for randomly placed volumes as for volumes centred on 
a particle (equation~\ref{borel}).  The same is true here.  
If ${\cal N}(j,b_1|k,b_2)$ denotes the average number of isolated 
$(j,b_1)$-volumes within a randomly placed $(k,b_2)$-volume, then 
\begin{displaymath}
F(j,b_1|k,b_2) = {\cal{N}}(j,b_1|k,b_2)\ {_1\!V_j\over _2\!V_k} = 
{\cal{N}}(j,b_1|k,b_2)\ {jb_1\over kb_2},
\end{displaymath}
so that 
\begin{equation}
{\cal{N}}(j,b_1|k,b_2) = {kb_2\over jb_1}\ F(j,b_1|k,b_2) 
= {k\over j}\ f(j,b_1|k,b_2).
\label{njk}
\end{equation}
The final expression is the same as equation~(45) of Sheth (1995).
Thus, equation~(\ref{njk}) shows that the average number of 
$(j,b_1)$-volumes that are within a $(k,b_2)$-volume is the same 
when $_2\!V_k$ is placed randomly in the Poisson distribution as 
when it is centred on a particle.  
In terms of collapsed haloes, this expression is similar to  
equation~(\ref{borel}), except that here the $(j,b_1)$-halo is 
constrained to be within a $(k,b_2)$-halo.  Thus, this expression 
gives the conditional mass function.  

Notice that 
\begin{equation}
{\cal{N}}(j,b_1|k,b_2)\to (k/j)\,f(j,b_1/b_2)\qquad {\rm when}\ k\gg j.
\end{equation}
Comparison with equation~(\ref{borel}) shows that, in this limit, 
the number density of $(j,b_1)$-volumes that are within a 
$(k,b_2)$-volume is the same as in the unconstrained case, the 
only difference is that $b\to b_1/b_2$, which reflects the 
fact that the background density within $_2\!V_k$ is 
$\bar n(1+\delta_2)$, rather than $\bar n$.  

All the arguments above were phrased entirely in terms of volumes 
that were concentric spheres.  This was done with a view to 
improving the clarity of the presentation---the entire analysis 
applies unchanged for arbitrarily shaped volumes.  This is because 
the underlying distribution is Poisson, so that all statements 
depend only on volumes $V$ and not their shapes, and all volumes 
can be broken up into mutually independent sub-volumes.  This is 
also why the dimensionality of the point distribution does not 
enter into the analysis anywhere.  
Appendix~\ref{bower} here shows this explicitly.  
In this respect, the statements above are obtained by an averaging 
process that is similar in spirit to that described in the Appendix 
of Bower (1991).  

\subsection{Clustering of haloes in Lagrangian space: the mean 
number of haloes}\label{fnhalo}
This section derives the first moment of the distribution of 
halo counts in randomly placed cells.  
The following sections describe the distribution of 
haloes in randomly placed cells when the halo mass is specified, 
and Section~\ref{range} considers the distribution for a 
range of masses.

To compute the mean number of haloes in randomly placed cells, 
it is useful to consider another way of computing $F(j,b_1)$.  
This alternative method also shows that dropping the Epstein (1983) 
and Sheth (1995) restriction (to only those volumes that are 
centred on particles) makes only a trivial difference to the 
final expression for $F(j,b_1)$.  

Let $f^{\rm I}(N,V_0)$ denote the probability that there are 
exactly $N$ particles within the sphere $V_0$, given that $V_0$ 
is centred on a randomly chosen particle in the Poisson 
distribution.  Then 
\begin{equation}
f^{\rm I}(N,V_0) = p(N-1,V_0),
\label{fint}
\end{equation}
where $p(k,V_0)$ is given by equation~(\ref{pois}).
The probability that there are $j$ particles in the sphere $V_j$ 
centered on the chosen particle, and all concentric spheres $V$ 
satisfying $V_j<V<V_0$ are less dense than $V_j$, given that there 
are $N$ particles in the concentric sphere $V_0>V_j$ is 
$f(j,b_1|N,b_0)$, where $1\le j\le N$, $b_1$ was defined above, 
and $b_0 \equiv \bar n V_0/N$.

Let $N_{01}$ denote the largest integer less than $\bar nV_0/b_1$; 
it is the maximum number of particles that may be in $V_0$, if $V_0$ 
is to be concentric to and less dense than $V_j$.  Also, let 
$f^{\rm e}(b_1|N,V_0)$ denote the probability that no sphere $V>V_0$ 
is concentric to and denser than the sphere $V_j$, given that there 
are exactly $1\le N\le N_{01}$ particles within $V_0$.  This 
quantity is just one minus the probability that there exists a 
sphere $V_k > V_0$ which is the largest sphere concentric to and 
having the same density as $V_j$, given that there are $N$ particles 
within $V_0$.  Then, 
\begin{equation}
f^{\rm e}(b_1|N,V_0) = 1 - \!\!\!
\sum_{k>N_{01}}^\infty \!\!\! p(k-N,V_k-V_0) f^{\rm e}(b_1|k,V_k),
\label{fenv}
\end{equation}
and equation~(\ref{fext}) shows that we can replace 
$f^{\rm e}(b_1|k,V_k)$ with $(1-b_1)$.  Define 
\begin{equation}
Q(b_1,N,V_0) \equiv p(N,V_0)\,f^{\rm e}(b_1|N,V_0) .
\label{qbig}
\end{equation}
Equations~(\ref{pjoint}), (\ref{fjb}) and~(\ref{fenv}) imply that  
\begin{equation}
Q(b_1,N,V_0) = 
p(N,V_0) - \sum_{k>N_{01}}^\infty \!\!\!p(N,V_0|k,V_k)\,F(k,b_2).   
\end{equation}
This, with equation~(\ref{pjfk}), shows that $Q=0$ when $N>N_{01}$.  

In terms of these quantities, 
\begin{equation}
f(j,b_1) = 
\sum_{N=j}^{N_{01}} f(j,b_1|N,b_0)\,f^{\rm I}(N,V_0)\,f^{\rm e}(b_1|N,V_0).
\label{fjv0}
\end{equation}
Now, equation~(\ref{vj}) implies that $b_0=\bar n V_0/N$, so this 
sum expresses $f(j,b_1)$ in terms of volumes $V_0$ that are certainly 
centred on a particle.  However, equations~(\ref{pois}) 
and~(\ref{fint}) show that 
\begin{displaymath}
f^{\rm I}(N,V_0) = p(N-1,V_0) = {N\over \bar nV_0}\,p(N,V_0) = p(N,V_0)/b_0,  
\end{displaymath}
so 
\begin{equation}
F(j,b_1) = \!\!
\sum_{N=j}^{N_{01}} \!\!F(j,b_1|N,b_0)\,p(N,V_0)\,f^{\rm e}(b_1|N,V_0) .
\label{fjmean}
\end{equation}
This final expression is written entirely in terms of randomly placed 
volumes, since $f^{\rm e}(b_1|N,V_0)$ depends only on the fact that 
there are exactly $N$ particles within $V_0$, and not on whether or 
not one of those particles is at the centre.  Straightforward but 
tedious algebra shows that this sum is consistent with the 
expressions for $f(j,b_1)$ and $F(j,b_1)$ derived earlier.  

This calculation can be easily manipulated to give the average 
number of isolated $(j,b_1)$-volumes that are in randomly placed 
cells of size $V_0$.  It is 
\begin{equation}
\bar n(j,b_1)V_0 \equiv 
\sum_{N=j}^{N_{01}} {\cal N}(j,b_1|N,b_0)\,Q(b_1,N,V_0) .
\label{nbarj}
\end{equation}
The sum on the right is $(\bar nV_0/jb_1)$ times the one in 
equation~(\ref{fjmean}), so it is equal to $\bar nV_0\,f(j,b_1)/j$.  
Comparison with equation~(\ref{borel}) shows explicitly that the 
mean number of isolated $(j,b_1)$-volumes that are in randomly 
placed cells of size $V_0$ is $V_0$ times the average density 
of these haloes, as required.  

\subsection{Cross correlation between haloes and mass}\label{laghm}
It is also straightforward to compute a measure of the cross 
correlation between $(j,b_1)$-haloes and the total number of particles 
that are in randomly placed cells of size $V_0$.  

Recall that ${\cal N}(j,b_1|N,b_0)$ denotes the average number
of $(j,b_1)$-haloes within an $(N,b_0)$-halo.  This expression 
also represents the average number of $(j,b_1)$ isolated regions that 
are within isolated regions $V_0$ which each have density $N/V_0$.  
Since these regions are isolated, they are different from a random 
region of size $V_0$ containing $N$ particles; recall that only a 
fraction $f^{\rm e}(b_1|N,V_0)$ of such random regions may contain 
a $b_1$-halo (and, of course, the number of particles in the 
$b_1$-halo may not exceed $N$).  The average number of 
$(j,b_1)$-haloes in the remaining $V_0$ cells (those that contain 
exactly $N>j$ particles and are not isolated) is zero. 

Thus, the average overabundance of $(j,b_1)$-haloes within the 
fraction $f^{\rm e}(b_1,N,V_0)$ of randomly placed $V_0$s that 
are isolated is 
\begin{equation}
\delta_{\rm h}^{\rm L}(j,b_1|N,b_0) = 
{{\cal N}(j,b_1|N,b_0)\over \bar n(j,b_1)V_0} - 1 
\label{bias}
\end{equation}
(Mo \& White 1996), and $\delta_{\rm h}^{\rm L} = -1$ in the 
remaining $V_0$s.  The superscript L represents the fact that 
this expression defines a bias relation that is associated with 
randomly placed regions $V_0$ in the initial Lagrangian space.  
As Mo \& White (1996) note, in general, dynamical evolution 
will result in a bias relation in Eulerian space that is different 
from this one in the Lagrangian space.  
Notice that, because $\delta_{\rm h}^{\rm L}$ is the average 
overabundance of haloes, it depends only on the first moment 
of the halo distribution.  To compute the rms scatter around this 
mean value requires knowledge of the higher order moments of 
the halo distribution.  We will compute this scatter later in 
this paper.  

When $N\gg j$, $f(j,b_1|N,b_0)\to f(j,b_1/b_0)$ (Appendix B in 
Sheth 1996), and $f^{\rm e}(b_1|N,V_0)\to 1$, so 
\begin{equation}
\delta_{\rm h}^{\rm L}(j,b_1|N,b_0) \to 
{N\over\bar nV_0}\,{f(j,b_1/b_0)\over f(j,b_1)} - 1 .
\label{bign}
\end{equation}
This relation will be useful later.  

Define 
\begin{equation}
\bar\xi_{\rm hm}^{\rm L}(j,b_1|V_0) \equiv
\left\langle \delta_{\rm h}^{\rm L}(1|0)\ \delta_0\right\rangle
\label{xihm}
\end{equation}
where $\delta_{\rm h}^{\rm L}(1|0)$ is given by equation~(\ref{bias}), 
\begin{displaymath}
\delta_0 = {N\over \bar nV_0} - 1,
\end{displaymath}
and the average above is over all randomly placed $V_0$.  
Writing all the terms out explicitly gives 
\begin{eqnarray}
\bar\xi_{\rm hm}^{\rm L}(j,b_1|V_0) &=& 
\left\langle{{\cal N}(1|0)\over\bar n(j,b_1)V_0}{N\over\bar nV_0}\right\rangle
- \left\langle{N\over \bar nV_0}\right\rangle \nonumber\\
&&\qquad\qquad -
\left\langle{{\cal N}(1|0)\over\bar n(j,b_1)V_0}\right\rangle + 1,
\end{eqnarray}
where $\bar n(j,b_1)$ is given by equation~(\ref{borel}), and 
${\cal N}(1|0)$ by equation~(\ref{njk}).  The second term in this 
expression is $(N/\bar nV_0)\,p(N,V_0)$, summed over all $N$, so 
it is unity, and it cancels the fourth term.  
The first and third terms have ${\cal N}(1|0)=0$ if they are not 
isolated, so they only recieve a non-zero contribution from the 
fraction $f^{\rm e}(b_1|N,V_0)$ of cells that are isolated.  
Writing the sum which gives the average explictly, 
and then using equation~(\ref{qbig}), shows that 
\begin{equation}
\bar\xi_{\rm hm}^{\rm L}(j,b_1|V_0) =
\sum_{N=j}^{N_{01}} \delta_0\ {{\cal N}(1|0)\over\bar n(j,b_1)V_0}\ 
Q(b_1,N,V_0).  
\label{xihmq}
\end{equation}
The upper limit on the sum comes from the fact that, if a randomly 
placed $V_0$ were to contain more particles, then it would be denser 
than $b_1$, so the $(j,b_1)$-regions inside it would not be 
isolated, and ${\cal N}(1|0)=0$.  This final expression is 
the cross correlation between $(j,b_1)$-haloes and mass, averaged 
over all randomly placed Lagrangian cells $V_0$.  

\subsection{The higher order moments of the halo distribution}\label{hom}
Previous subsections computed the mean number of isolated 
$(j,b_1)$-regions, i.e., the mean number of $(j,b_1)$-haloes that 
are in randomly placed cells of size $V_0$.  This subsection computes 
the higher order moments of the distribution.  To do so, it is 
necessary to examine the expression for ${\cal N}(j,b_1|N,b_0)$ in 
more detail.  

Let $p({\bmath n},b_1|N,b_0)$, where ${\bmath n}=(n_1,\cdots ,n_N)$ 
and $b_0\ge b_1$, denote the probability that the volume 
$V_0= {_0\!V_N}$ is composed of $m$ isolated subvolumes, 
of which there are $n_j$ isolated $(j,b_1)$-volumes (each of size 
$_1\!V_j$), and $1\le j\le N$.  Thus, $\sum_{j=1}^N n_j = m$, and 
mass conservation requires that $\sum_{j=1}^N j\,n_j = N$.  
Sheth (1996) describes a model, based on the Poisson distribution, 
in which 
\begin{equation}
p({\bmath n},b_1|N,b_0) =
{(Nb_{01})^{m-1}{\rm e}^{-Nb_{01}}\over \eta(N,b_0)}\,
\prod_{j=1}^N {\eta(j,b_1)^{n_j}\over n_j!},
\label{prtfnc}
\end{equation}
where $b_{01}=(b_0-b_1)$, and $Nb_0 = \bar nV_0$.  
See Sheth \& Pitman (1997), and Sheth \& Lemson (1998), 
for other interpretations of this partition formula.  

For this model, the average number of isolated regions containing 
exactly $j$ particles, each with average density parametrized by 
$b_1$, that are within spheres of size $V_0$ containing exactly 
$N$ particles is given by 
\begin{eqnarray}
\langle n_j,b_1|N,b_0\rangle &=& 
\sum_{\pi[{\bmath n}]} n_j\ p({\bmath n},b_1|N,b_0) \nonumber \\
&=& {N\over j}\,f(j,b_1|N,b_0) = {\cal N}(j,b_1|N,b_0)  ,
\label{meanj}
\end{eqnarray}  
where $\pi[{\bmath n}]$ denotes the set of all distinct ordered 
partitions of $N$ (Appendix~B in Sheth 1996).  

To set notation, it is useful to rewrite some of the expressions 
derived earlier.  Let $n_j(b_1,{\bmath n},N,b_0)$ denote 
the number of $(j,b_1)$-haloes in the partition ${\bmath n}$ of $N$.  
(In the formula above, this was simply written as $n_j$.)  
Then 
\begin{displaymath}
{\cal N}(j,b_1|N,b_0) \equiv \sum_{\pi[{\bmath n}]}
n_j(b_1,{\bmath n},N,b_0)\ p({\bmath n},b_1|N,b_0).  
\end{displaymath}
Define 
\begin{equation}
\Delta_j(b_1,{\bmath n},N,b_0) \equiv 
{n_j(b_1,{\bmath n},N,b_0)\over \bar n(j,b_1)V_0} - 1 ,
\end{equation}
where $Nb_0\equiv \bar nV_0$.  
This is the overdensity of $(j,b_1)$-haloes in the partition 
${\bmath n}$ of $N$, relative to the average density of such 
haloes.  This, averaged over all partitions, gives the average bias 
relation of equation~(\ref{bias}):
\begin{equation}
\delta_{\rm h}^{\rm L}(j,b_1|N,b_0) \equiv 
\sum_{\pi[{\bmath n}]} \Delta_j(b_1,{\bmath n},N,b_0)\ 
p({\bmath n},b_1|N,b_0) .
\end{equation}
The variance in this bias relation is  
\begin{equation}
{\rm Var}(\Delta_j) = 
\Bigl\langle \Delta_j^2(b_1,{\bmath n},N,b_0)\Bigr\rangle - 
\Bigl\langle\Delta_j(b_1,{\bmath n},N,b_0)\Bigr\rangle^2 ,
\end{equation}
where the average is over all partitions ${\bmath n}$ of $N$.  
This is the same as 
\begin{displaymath}
{\rm Var}(\Delta_j) = 
{\Bigl\langle n_j^2(b_1,{\bmath n},N,b_0)\Bigr\rangle
\over [\bar n(j,b_1)V_0]^2} - 
{\Bigl\langle n_j(b_1,{\bmath n},N,b_0)\Bigr\rangle^2
\over [\bar n(j,b_1)V_0]^2} ,
\end{displaymath}
where the averages are over all partitions ${\bmath n}$ of $N$.  
The first term is the second moment of the distribution of 
$(j,b_1)$-subhaloes within $(N,b_0)$-haloes.  
The rms scatter around the mean bias relation is the square root 
of Var$(\Delta_j)$.  So, to compute the 
scatter in the bias relation requires knowledge of the second 
moment of the halo distribution.  Fortunately, for the model 
described by equation~(\ref{prtfnc}), all such higher order moments 
are known.  

The factorial moment of order $\alpha$, of the distribution of 
$(j,b_1)$-haloes within $(N,b_0)$-haloes, is 
\begin{eqnarray}
\mu_\alpha(j,b_1|N,b_0)&\equiv&  
\left\langle\!{n_j!\over (n_j-\alpha)!},b_1\,\Biggl\vert\,N,b_0\!\right\rangle 
\nonumber \\
&=&\Bigl[N(b_0-b_1)\Bigr]^\alpha\ 
{\eta^\alpha(j,b_1)\ \eta(m,B)\over\eta(N,b_0)} 
\label{njfac}
\end{eqnarray}
where 
\begin{equation}
mB \equiv (N-\alpha j)B = Nb_2-\alpha jb_1 
\end{equation}
(Appendix~B of Sheth 1996).  
Similarly, cross--moments are given by 
\begin{eqnarray}
&&\!\!\!\!\!\!\!\!\!\!\!\!
\left\langle\!{n_i!\over (n_i-\alpha)!}{n_j!\over
(n_j-\beta)!},b_1\,\Biggl\vert\,N,b_0\!\right\rangle \nonumber \\
&&\ =\ \Bigl[N(b_0-b_1)\Bigr]^{\alpha + \beta} \ 
{\eta^\alpha(i,b_1)\,\eta^\beta(j,b_1)\ \eta(m,B)\over\eta(N,b_0)} ,
\label{facij}
\end{eqnarray} 
where 
\begin{equation}
mB\equiv (N-\alpha i- \beta j)B = Nb_0-\alpha ib_1 - \beta jb_1 .
\end{equation}

These formulae for the higher order moments were obtained after 
using equation~(\ref{prtfnc}) for the partition formula.  
Sheth \& Lemson (1998) show that this formula arises naturally 
as a consequence of the fact that disconnected volumes in a Poisson 
distribution are mutually independent.  This allows a simple 
interpretation of equation~(\ref{facij}).  

Define 
\begin{eqnarray}
c(i,j,b_1|k,b_0) &\equiv&
{\cal N}(j,b_1|k,b_0)\ {\cal N}(i,b_1|k-j,b') \nonumber \\
&=& \langle n_j,b_1|k,b_0\rangle\ \langle n_i,b_1|k-j,b'\rangle ,
\label{cijk}
\end{eqnarray}
where 
\begin{equation}
{\bar n\over b'} \equiv 
{k-j\over _0\!V_k -\, _1\!V_j} \equiv \bar n(1+\delta').
\label{bprime}
\end{equation}  
The halo containing $j$ particles can be thought of as occupying 
$_1\!V_j$ of the total volume $_0\!V_k$.  Thus, $b'$ parametrizes 
the density in the remaining volume $_0\!V_k -\, _1\!V_j$, which 
contains $(k-j)$ particles.  Thus, $c(ij|k)$ is the product of 
the mean number of $(j,b_1)$-haloes within the volume associated 
with the $(k,b_0)$-halo and the mean number of $(i,b_1)$-haloes 
in the remaining volume, given that there is a $(j,b_1)$-halo 
within $_0\!V_k$.  
Now, equation~(\ref{meanj}), implies that 
\begin{eqnarray}
c(i,j,b_1|k,b_0) &=& {k\over j}\,f(j,b_1|k,b_0)\ 
{(k-j)\over i}\,f(i,b_1|k-j,b')\nonumber \\
&=& k(b_0-b_1)\,{\eta(j,b_1)\,\eta(k-j,b')\over \eta(k,b_0)} 
\nonumber\\
&&\times\ 
(k-j)(b'-b_1)\nonumber \\
&&\times\ {\eta(i,b_1)\,\eta(k-j-i,b'')\over \eta(k-j,b')} ,
\end{eqnarray}
where $b''$ is defined similarly to $b'$.  That is, 
\begin{equation}
{\bar n\over b''} = {k-j-i\over _0\!V_k -\, _1\!V_j -\, _1\!V_i}.
\end{equation}
However, 
\begin{equation}
(k-j)\,(b'-b_1) =  (kb_0-jb_1) - (k-j)b_1 = k(b_0-b_1) ,
\label{kb21}
\end{equation}
so that 
\begin{displaymath}
c(i,j,b_1|k,b_0) = 
{[k(b_0-b_1)]^2\over \eta(k,b_0)}\,
\eta(j,b_1)\eta(i,b_1)\eta(k-j-i,b'').
\end{displaymath}
This expression is symmetric in $i$ and $j$, and it is easy to 
see that it is the same as 
\begin{equation}
c(i,j,b_1|k,b_0) = 
{\cal N}(i,b_1|k,b_0)\ {\cal N}(j,b_1|k-i,b'),
\end{equation}
with the appropriate redefinition of $b'$.  
Simple algebra shows that 
\begin{equation}
c(i,j,b_1|k,b_0) = 
\bigl\langle n_i\,n_j,b_1|k,b_0\bigr\rangle ,
\label{cij}
\end{equation}
where the right hand side is equation~(\ref{facij}) with 
$\alpha=\beta=1$.  This shows explicitly that 
\begin{equation}
\bigl\langle n_i\,n_j,b_1|k,b_0\bigr\rangle
= \bigl\langle n_j,b_1|k,b_0\bigr\rangle\ 
\left\langle n_i,b_1|k-j,b'\right\rangle ,
\end{equation}
and that it was obtained by treating the volumes $_1\!V_j$ 
and $_0\!V_k - _1\!V_j$ as being disconnected from, and 
independent of, each other.  

This argument can be generalized to the higher order moments.  
For example, if 
\begin{equation}
(k - n j)\,b^{(n)} = kb_0 - n j b_1, \qquad{\rm with}\ b^{(0)}=b_0,
\end{equation}
then 
\begin{equation}
(k- n j)\,(b^{(n)} - b_1) = k(b_0-b_1).  
\end{equation}
So equation~(\ref{njfac}) is also equal to 
\begin{eqnarray}
\mu_\alpha(j,b_1|N,b_0) &=& 
\prod_{n=0}^{\alpha-1}\ \Bigl\langle n_j,b_1|k-nj,b^{(n)}\Bigr\rangle
\nonumber \\
&=& \prod_{n=0}^{\alpha-1}\ {\cal N}\Bigl(j,b_1|k-nj,b^{(n)}\Bigr).
\label{musl98}
\end{eqnarray}
Thus, the higher order moments described by equation~(\ref{njfac}) 
are consistent with the fact that disconnected volumes in a Poisson 
distribution are mutually independent.  The cross correlation 
moments of equation~(\ref{facij}) can be interpretted similarly.  
Thus, for example, the variance in the bias relation above is 
\begin{eqnarray}
{\rm Var}(\Delta_j) &=& 
{c(j,j,b_1|N,b_0)+{\cal N}(j,b_1|N,b_0)\over [\bar n(j,b_1)V_0]^2} 
\nonumber \\
&&\qquad - {{\cal N}(j,b_1|N,b_0)^2\over [\bar n(j,b_1)V_0]^2}.
\end{eqnarray}

Equation~(\ref{meanj}) in~(\ref{nbarj}) implies that 
\begin{equation}
\bar n(j,b_1)V_0 = 
\sum_{N=j}^{N_{01}}\, \langle n_j,b_1|N,b_0\rangle\ Q(b_1,N,V_0) .
\label{mnsum}
\end{equation}  
This shows how the average number of isolated regions, each with 
average internal density $\bar n(1+\delta_1)$ and each containing 
$j$ particles, that are within randomly placed volumes $V_0$, can 
be obtained from the partition formula of equation~(\ref{prtfnc}).  
The main reason for writing this expression explicitly is that 
it shows clearly how to compute the higher order moments associated 
with the model of equation~(\ref{prtfnc}).  

Let $M_\alpha(j,b_1)$ denote the $\alpha$th factorial moment of the 
distribution of $(j,b_1)$-regions that are within spheres 
of size $V_0$.  It is obtained by a similar average to that for 
the mean:  
\begin{equation}
M_\alpha(j,b_1|V_0) =\sum_{N=\alpha j}^{N_{01}}
\mu_\alpha(j,b_1|N,b_0)\ Q(b_1,N,V_0).
\label{muijk}
\end{equation}
When $\alpha=1$, this is the same as equation~(\ref{mnsum}).  

Let $\bar\xi_{\rm hh}^{\rm L}(ij|0)$ denote the correlation 
between isolated $(i,b_1)$- and $(j,b_1)$-regions, averaged over 
Lagrangian cells of size $V_0$.  Then 
\begin{equation}
M_2(j,b_1|V_0) = \Bigl(\bar n(j,b_1)V_0\Bigr)^2
\Bigl(1 + \bar\xi_{\rm hh}^{\rm L}(jj|0)\Bigr).  
\end{equation}
Similarly, when the $b_1$-isolated regions do not have the same 
number of particles, 
\begin{equation}
1+\bar\xi_{\rm hh}^{\rm L}(ij|0) = \!\!\sum_{N=i+j}^{N_{01}}
{c(i,j,b_1|N,b_0)\over \bar n(i,b_1)V_0\,\bar n(j,b_1)V_0}\,
Q(b_1,N,V_0) ,
\label{xiijv0}
\end{equation}
with the understanding that $c(ij|N)=0$ if $(i+j)>N$, so that 
$\bar\xi_{\rm hh}^{\rm L}(ij|0)=-1$ if $(i+j)>N_{01}$.  

Suppose that each isolated $b_1$-region within $V_0$ is
represented by (a randomly chosen) one of its constituent 
particles.  This defines a point process, for which statistics 
such as the distribution of halo counts-in-cells can be computed.  
Since $(j,b_1)$-regions are associated with $(j,b_1)$-haloes, 
it is convenient to call the randomly chosen representative point 
of such a halo its centre-of-mass.  The expressions above 
give the higher order moments of the distribution of counts of 
haloes in randomly placed cells $V_0$.  Halo--halo correlations 
can be computed from these moments.  For example, 
equation~(\ref{xiijv0}) gives the volume averaged correlation 
function of $(i,b_1)$- and $(j,b_1)$-haloes.  All the necessary 
sums can be evaluated analytically.  

\subsection{Statistics for a range of halo masses}\label{range}
The previous subsections considered the halo distribution when 
the halo mass was specified.  This subsection shows how to 
compute correlations between haloes that have a range of 
masses.  This is necessary, since comparison with simulations 
is typically done by considering averages over a range of 
masses, and, as we discuss below, the transition to considering 
a range of masses is not completely straightforward.  That is, 
simply integrating the previous expressions over the relevant mass 
range, weighted by the unconditional mass function, is not 
entirely correct.  It turns out that, over a large range of 
scales, the correct expression yields only a minor correction 
to the naive expression, so readers interested only in results 
may prefer to skip this section.   

So far, the distribution of isolated regions and that of the 
centre-of-mass distribution of collapsed haloes were assumed to 
be the same.  However, there is an important difference between 
haloes and isolated regions.  Namely, by definition, a Lagrangian 
volume $V_0$ with overdensity $\delta_0$ cannot contain an isolated 
$V_1<V_0$ region of overdensity $\delta_1<\delta_0$, nor can it 
contain an isolated region of density $\delta_1\le\delta_0$ if its 
size is $V_1>V_0$.  Thus, the number of such isolated regions within 
an overdense or non-isolated cell $V_0$ is zero.  

However, since a collapsed halo is represented only by the volume 
element associated with its centre-of-mass, haloes are said to lie 
within a cell if their centre-of-mass does.  Thus, an $M_1$ halo 
may well lie within a $V_0$ cell, even if $M_1>M_0$.  
Moreover, in the model, a region $V_0$ of density 
$\delta_0\ge \delta_1$ is certainly a subregion of an isolated 
$\delta_1$ region, with $V_1>V_0$.  Such an overdense $V_0$ is 
said to contain the $M_1$ halo only if the volume element that 
represents the centre-of-mass of the halo falls inside it; in the 
model, the centre-of-mass is a randomly chosen volume element, so 
this happens with probability $(V_0/V_1)$.  Thus, a cell $V_0$ 
that is either overdense or not isolated may contain a halo, whereas, 
by definition, it cannot contain an isolated region.  
Previously, this difference between haloes and isolated regions 
was unimportant.  Now, however, since we must integrate over 
a range of halo masses, it can be important.  

Consider the set of $V_0$ cells placed randomly in the 
Lagrangian space.  
Suppose we wish to count up the number of $b_1$-haloes that 
are more massive than $m$, that are in such cells.  Given a 
value of $b_1$, these cells can be divided into two classes:  
those that are isolated and those that are not.  Those that 
are isolated can be classified by the number $N$ of particles 
within the cell.  All isolated cells that contain $N$ particles 
can be further classified by the way in which the $N$ particles 
are divided into $b_1$-haloes.  
Consider an isolated cell $V_0$ that is known to contain 
exactly $N$ particles which are partitioned into $b_1$-haloes.  
As before, denote the particular partition by the vector 
${\bmath n}$.  Let $N_{\rm h}(j,b_1|{\bmath n},b_0)$ denote the number 
of $(j,b_1)$-haloes that are within such a cell.  The number of 
$b_1$-haloes more massive than $m$ that are within such cells is 
\begin{equation}
N_{\rm isol}(>\!m,b_1|{\bmath n},V_0) = 
\sum_{j>m}^N N_{\rm h}(j,b_1|{\bmath n},b_0).
\end{equation}
Equation~(\ref{meanj}) shows that this quantity, averaged over 
all partitions of $N$, is 
\begin{displaymath}
\sum_{\pi[{\bmath n}]} N_{\rm isol}(>\!m,b_1|{\bmath n},V_0)\ 
p({\bmath n};b_1|N,b_0) = 
\sum_{j>m}^N \bigl\langle n_j,b_1|N,b_0\bigr\rangle.
\end{displaymath}
This, averaged over all values of $N$, is 
\begin{equation}
\bar N_{\rm isol}(>\!m,b_1|V_0) = \sum_N Q(b_1,N,V_0)
\sum_{j>m}^N \bigl\langle n_j,b_1|N,b_0\bigr\rangle ,
\label{nisoprt}
\end{equation}
since $Q(b_1,N,V_0)$ denotes the fraction of the total number of 
cells that are isolated.  This sum is zero when $N\le m$, because 
if the cell $V_0$ is isolated, then all the particles associated 
with a halo within $V_0$ must be contained in $V_0$, and we are 
only counting haloes more massive than $m$.  The definition of 
$Q$ (equation~\ref{qbig}) insures that the sum is also zero when 
$N>N_{01}$.  This is because, when $N>N_{01}$, then the cell is denser 
than $b_1$, so it is not isolated on the scale $V_0$.  The order of 
the sums above can be interchanged to yield 
\begin{eqnarray}
\bar N_{\rm isol}(>\!m,b_1|V_0) &=& 
\sum_{j>m}^{N_{01}} \sum_{N=j}^{N_{01}} 
Q(b_1,N,V_0)\ {\cal N}(j,b_1|N,b_0) \nonumber \\
&=& \sum_{j>m}^{N_{01}}  \bar n(j,b_1)V_0 , 
\label{nisol}
\end{eqnarray}
where the final equality follows from equation~(\ref{nbarj}).  

Cells that are not isolated on scale $V_0$ can be classified by 
the scale $_1\!V_j>V_0$ at which they first become isolated.  
They can be further classified by the number of particles $N<j$ 
they actually contain on scale $V_0$.  The probability that a cell 
first becomes isolated on scale $_1\!V_j$, given that it contains 
$N$ particles on scale $V_0<{_1\!V_j}$ is 
\begin{equation}
P(j,b_1|N,V_0) \equiv {p(N,V_0|j,{_1\!V_j})\ F(j,b_1)\over p(N,V_0)}.
\end{equation}
Recall that $F(j,b_1)$ is the probability that a randomly 
placed cell is isolated on the scale $_1\!V_j=jb_1/\bar n$, so 
the expression above follows from Bayes' rule.  
The region $V_0$ is a subregion within the isolated region 
$_1\!V_j$.  Since $_1\!V_j$ is isolated, it can be thought of as 
a $(j,b_1)$-halo.  The subregion $V_0$ is said to contain this 
$(j,b_1)$-halo only if it contains the randomly chosen 
centre-of-mass particle of the halo.  This happens with 
probability $V_0/{_1\!V_j}$.  Therefore, the average number of 
$b_1$-haloes that are in cells which are not isolated on scale 
$V_0$ is 
\begin{displaymath}
\bar N_{\rm other}(>\!m,b_1|V_0) =
\end{displaymath}
\begin{equation}
\sum_{N=0}^\infty p(N,V_0)\!\sum_{j=j_{\rm min}}^\infty
{V_0\over _1\!V_j}\, P(j,b_1|N,V_0) ,
\label{nother}
\end{equation}
where $j_{\rm min}=(m+1)$ if $m>N_{01}$.  Otherwise,  
$j_{\rm min}=(N_{01}+1)$.  
Since $V_0<{_1\!V_j}$, $p(N,V_0|j,{_1\!V_j})=0$ if $N>j$.  
With this in mind, the order of the sums can be interchanged:  
\begin{displaymath}
\bar N_{\rm other} = 
\sum_{j=j_{\rm min}}^\infty (V_0/{_1\!V_j})\,F(j,b_1) 
\sum_{N=0}^j p(N,V_0|j,{_1\!V_j}).
\end{displaymath}
The sum over $N$ is unity, so the average number of $b_1$-haloes 
more massive than $m$ that are within such $V_0$ cells is 
\begin{equation}
\bar N_{\rm other} = 
\sum_{j=j_{\rm min}}^\infty {V_0\over{_1\!V_j}}\,F(j,b_1) 
= \sum_{j=j_{\rm min}}^\infty \bar n(j,b_1)V_0 .
\end{equation}
The final equality follows from equation~(\ref{borel}).  

On average, the number of $b_1$-haloes that are more massive 
than $m$, that are within randomly placed $V_0$ cells, is given 
by adding the contribution from the two types of cells---those that 
are isolated on scale $V_0$ and those that are not.  
Thus, when $m<N_{01}$, then the average over all $V_0$ cells, 
$\bar N_{\rm isol} + \bar N_{\rm other}$, is 
\begin{equation}
\bar n(>\!m,b_1)V_0 \equiv \sum_{j>m}^\infty \bar n(j,b_1)V_0 .
\end{equation}
If $m>N_{01}$, then $\bar N_{\rm isol}(>\!m,b_1|V_0)=0$, and 
the average is simply $\bar N_{\rm other}(>\!m,b_1|V_0)$ which 
is the same as the expression above.  

As before, define
\begin{equation}
\Delta_{\rm h}(>\!m,b_1|{\bmath n},V_0) \equiv
{N_{\rm h}(>\!m,b_1|{\bmath n},V_0)\over \bar n(>\!m,b_1)V_0} - 1.
\end{equation}
The cross correlation between haloes and mass, averaged over 
all cells $V_0$, is 
\begin{eqnarray}
\bar\xi_{\rm hm}^{\rm L}(>\!m,b_1|V_0) &\equiv&
\bigl\langle \Delta_{\rm h}(>\!m,b_1|{\bmath n},V_0)\ \delta_0\bigr\rangle 
\nonumber \\
&=& {\bigl\langle N_{\rm h}(>\!m,b_1|{\bmath n},V_0)\ \delta_0\bigr\rangle
\over \bar n(>\!m,b_1)V_0} ,
\end{eqnarray}
since $\delta_0=(N-\bar N_0)/\bar N_0$.  

For isolated cells, this average can be computed in two steps.  
The first is to average over all partitions $\pi[{\bmath n}]$ of $N$. 
The second is to average over all values of $N$.  
If $m\le N_{01}$, then the contribution from isolated cells is 
\begin{eqnarray}
\bigl\langle N_{\rm isol}\ \delta_0\bigr\rangle &=& 
\sum_{j>m}^{N_{01}} \sum_{N=j}^{N_{01}} 
\delta_0\ Q(b_1,N,V_0)\ {\cal N}(j,b_1|N,b_0)\nonumber \\
&=& \sum_{j>m}^{N_{01}} \bar n(j,b_1)V_0\ \bar\xi_{\rm hm}^{\rm L}(j,b_1|V_0),
\end{eqnarray}
where the first equality arises from the average over partititions 
of $N$ (equation~\ref{nisoprt}), and the second equality follows 
from using equation~(\ref{xihmq}).
The contribution from the other cells is 
\begin{equation}
\bigl\langle N_{\rm other}\ \delta_0\bigr\rangle = 
\sum_{j>N_{01}}^\infty {V_0\over _1\!V_j} F(j,b_1)
\sum_{N=0}^j \delta_0\ p(N,V_0|j,{_1\!V_j}).
\end{equation}
Since $\delta_0 = (N-\bar N)/\bar N_0$, the sum over $N$ is 
\begin{displaymath}
{j\over \bar N_0}{V_0\over _1\!V_j} - 1 = {1-b_1\over b_1} 
= \delta_1,
\end{displaymath}
so the contribution from these other cells is 
\begin{equation}
\bigl\langle N_{\rm other}\ \delta_0\bigr\rangle \ = 
\ \delta_1\!\sum_{j>N_{01}}^\infty \bar n(j,b_1) V_0.
\end{equation}
The cross correlation function averaged over all cells is the 
sum of these two terms divided by $\bar n(>\!m,b_1)V_0$:  
\begin{eqnarray}
\bar\xi_{\rm hm}^{\rm L}(>\!m,b_1|V_0) &=& 
\sum_{j>m}^{N_{01}} {\bar n(j,b_1)V_0\ 
\bar\xi_{\rm hm}^{\rm L}(j,b_1|V_0)\over \bar n(>\!m,b_1)V_0}\nonumber \\
&& \ \ \ +\ \ \delta_1\!\sum_{j>N_{01}}^\infty 
{\bar n(j,b_1) V_0\over \bar n(>\!m,b_1)V_0}.
\label{rangehm}
\end{eqnarray}
There is no contribution from isolated cells, and the remaining 
cells yield 
\begin{displaymath}
\bar\xi_{\rm hm}^{\rm L}(>\!m,b_1|V_0) = \delta_1,\qquad{\rm if}\ m>N_{01} .
\end{displaymath}

Auto-correlations between haloes can be computed similarly.  
Define 
\begin{equation}
\bar\xi_{\rm hh}^{\rm L}(>\!m,b_1|V_0) \equiv 
\bigl\langle \Delta^2_{\rm h}(>\!m,b_1|{\bmath n},V_0)\bigr\rangle - 
{1\over \bar n(>\!m,b_1)V_0}.
\end{equation}
The second term is the shot-noise term. It accounts for the 
fact that the halo distribution is discrete.  

First consider the case when $m\le N_{01}$, so isolated 
cells may contain more than one halo in the mass range of 
interest.  
For isolated cells, correlations arise as a result of two averages.  
The first is over all partitions of $N$.  The second is over all 
values of $N$.  Given a partition ${\bmath n}$ of $N$, 
\begin{displaymath}
N^2_{\rm isol} = (n_{m+1} + \cdots +n_N)^2 
= \sum_{i>m}^N \sum_{j>m}^{N} n_i\,n_j.
\end{displaymath}
Equations~(\ref{njfac}) and~(\ref{facij}) show how to compute
these averages over the set of partitions $\pi[{\bmath n}]$.  
Notice that when $i=j$, then~(\ref{facij}) for 
$\langle n_i\,n_j\rangle$, is the same as~(\ref{njfac}) for 
$\langle n_i\,(n_i-1)\rangle$.  Therefore, if we use~(\ref{facij})
even when $i=j$, and write it using~(\ref{cij}), then the average 
over $N$ is 
\begin{eqnarray}
\left\langle N^2_{\rm isol}\right\rangle &=& 
\sum_{N} Q(b_1,N,V_0) \sum_{i>m}^{N_{01}} \sum_{j>m}^{N_{01}}
c(i,j,b_1|N,b_0) \nonumber \\
&&+\ \sum_{N} Q(b_1,N,V_0)\sum_{j>m}^{N_{01}}
\bigl\langle n_j,b_1|N,b_0\bigr\rangle
\label{nhsq}
\end{eqnarray}
where $c(ij|N) = \langle n_i\,n_j|N\rangle = 0$ if $(i+j)>N$.  
Equation~(\ref{nisol}) shows that the second term is just 
$\bar N_{\rm isol}(>\!m,b_1|V_0)$.  

Cells $V_0$ that are not isolated either contain one or no 
haloes.  So, the contribution from these cells is just 
$\bar N_{\rm other}(>\!m,b_1|V_0)$ of equation~(\ref{nother}).  
The contribution from these cells, plus the second term from 
the isolated cells equals $\bar n(>\!m,b_1)V_0$.  
Together, they cancel the shot noise term in the definition of 
$\bar\xi_{\rm hh}^{\rm L}$.  The order of the sums in the remaining 
first term of~(\ref{nhsq}) can be rearranged to yield 
\begin{displaymath}
1+\bar\xi_{\rm hh}^{\rm L}(>\!m,b_1|V_0) = 
\sum_{i>m}^{N_{01}} \sum_{j>m}^{N_{01}} 
{\bar n(i,b_1)V_0\ \bar n(j,b_1)V_0\over\bar n^2(>\!m,b_1)V_0^2}
\end{displaymath}
\begin{equation}
\qquad\qquad \times \ \sum_{N=i+j}^{N_{01}} 
{c(i,j,b_1|N,b_0)\over \bar n(i,b_1)V_0\,\bar n(j,b_1)V_0}
\ Q(b_1,N,V_0) ,
\end{equation}
where $c(i,j|N)=0$ if $(i+j)>N$.  
If $m>N_{01}$, there are no isolated cells which contain 
haloes in the mass range of interest.  All other cells either 
contain one or no haloes, so, for these cells 
$N^2_{\rm h}(>\!m,b_1|V_0) = \bar n(>\!m,b_1)V_0$.  This term 
cancels the shot noise term, so that 
\begin{equation}
\bar\xi_{\rm hh}^{\rm L}(>\!m,b_1|V_0) = -1,\qquad{\rm if}\ m>N_{01}.
\end{equation}
Comparison with equation~(\ref{xiijv0}) shows that  
\begin{eqnarray}
1+\bar\xi_{\rm hh}^{\rm L}(>\!m,b_1|V_0) &=& 
\sum_{i>m}^{N_{01}} \sum_{j>m}^{N_{01}} 
{\bar n(i,b_1)V_0\ \bar n(j,b_1)V_0\over\bar n^2(>\!m,b_1)V_0^2}
\nonumber \\
&&\ \ \times\ \bigl[1+\bar\xi_{\rm hh}^{\rm L}(i,j,b_1|V_0)\bigr],
\label{rangehh}
\end{eqnarray}
with  the convention that $1+\bar\xi_{\rm hh}^{\rm L}(ij|0)=0$ 
if $(i+j) > N_{01}$.  

\subsection{Clustering from white noise as a limit of the Poisson model}\label{wnlim}

This subsection shows explicitly that, in the limit of small 
fluctuations and large numbers of particles, all the statements about 
clustering from white noise initial conditions presented in the 
main text can be derived from the Poisson statements derived above 
by using Stirling's approximation for all the factorials, and 
writing all expressions to lowest order in $\delta$.  

Let $N=\bar nV(1+\delta)$, and let 
$S\equiv\langle\delta^2\rangle = 1/\bar nV$ denote the mean 
square fluctuation of $\delta = (N-\bar nV)/\bar nV$ in cells of 
size $V$.  Then ${\rm d}N/{\rm d}\delta = \bar nV$, and, 
when $\delta\ll 1$, then use of Stirling's approximation 
for the factorial reduces equation~(\ref{pois}) for $p(M,V)$
to equation~(\ref{g0}).  
Similarly, equation~(\ref{pjoint}) tends to equation~(\ref{g10}).  
Furthermore, 
$f(M,b) \to f(S,\delta)\,|{\rm d}S/{\rm d}M|\,{\rm d}M$
of equation~(\ref{fm}) (e.g. Epstein 1983) and 
$f(M_1,b_1|M_0,b_0) \to 
f(S_1,\delta_1|S_0,\delta_0)\,|{\rm d}S_1/{\rm d}M_1|\,{\rm d}M_1$
of equation~(\ref{fm1m2}) (Sheth 1995), 
since $b=1/(1+\delta)\approx (1 - \delta)$, and 
$1 - (b_1/b_2) \approx (\delta_1-\delta_2)$.  
Simple algebra shows that equations~(\ref{g10}) and~(\ref{fm1m2}) 
satisfy a recursion relation that is similar to the one in the 
discrete Poisson case (and solved in Appendix~\ref{solvrec}).  
Namely, 
\begin{equation}
p(S_1,\delta_1|S_0,\delta_0) =\! 
\int_{S_0}^{S_1} \!\!\!\! f(S',\delta_1|S_0,\delta_0)\, 
p(S_1,\delta_1|S',\delta_1)\ {\rm d}S' .
\label{grecur}
\end{equation}
By considering the statistics of trajectories that are analogous 
to those considered in the Poisson case, Bond et al. (1991) have 
shown that these expressions can be derived directly from the 
white noise field itself.  

The virtue of using the trajectory description is that it 
allows one to see the correctness of many statements that are 
otherwise tedious to compute.  For example, suppose we label each 
trajectory by the value $S'$, which is the smallest value of $S$ 
at which it has overdensity density $\delta'$.  
If $\delta\ge \delta'\ge 0$, then 
\begin{equation}
p(S,\delta) = \int_0^S p(S,\delta|S',\delta')\,f(S',\delta')\ {\rm d}S'.
\label{pdfs}
\end{equation}
The left hand side of this expression is the set of all trajectories 
that pass through $\delta$ at $S$.  The right hand side is the 
set of all trajectories that first pass through $\delta'\le \delta$ 
at $S'<S$, and then pass through $\delta$ at $S$, summed over all 
$S'\le S$, since trajectories that first pass through $\delta'$ 
on scale $S'>S$ have certainly not passed through $\delta\ge \delta'$ 
at $S$.  Clearly, the left hand side equals the right.  
When $\delta=\delta'$, then direct substitution shows that this 
is correct.  Otherwise, direct substitution is not the easiest way 
to see that this must be correct.  This equation is the analogue 
of equation~(\ref{pjfk}).  

Notice that
\begin{equation}
{{\rm d}\over {\rm d}\delta}\,p(S,\delta) = f(S,\delta),
\label{psfs}
\end{equation}
and 
\begin{equation}
{{\rm d}\over {\rm d}\delta}\,p(S,\delta|S',\delta') = 
f(S,\delta|S',\delta').
\end{equation}
These relations, with equation~(\ref{pdfs}), imply that 
\begin{eqnarray}
{{\rm d}\,p(S,\delta)\over {\rm d}\,\delta} &=& 
{{\rm d}\over {\rm d}\delta}
\int_0^S p(S,\delta|S',\delta')\,f(S',\delta')\ {\rm d}S' \nonumber \\
&=& \int_0^S f(S,\delta|S',\delta')\,f(S',\delta')\ {\rm d}S'
\nonumber \\
&=& f(S,\delta)
\end{eqnarray}
as required by equation~(\ref{psfs}).  

The number density of $M_1$ haloes, that is, the unconstrained mass 
function, is $\bar\rho f(M_1,\delta_1)/M_1$ which is the same 
as equation~(\ref{nm1}).  
Similarly, the conditional mass distribution is 
$(M_0/M_1)\,f(M_1,\delta_1|M_0,\delta_0)$ which is the 
same as equation~(\ref{n10}).  
These are the analogues of equations~(\ref{borel}) and~(\ref{njk}).  

The limit of equation~(\ref{qbig}) is 
\begin{eqnarray}
&&\!\!\!\!\!\!\!\!\!\!\!\!\!\!\!\! 
Q(b_1,M_0,V_0)\to q(\delta_1,\delta_0,V_0) \nonumber \\
&&\!\!\!\!\!\!\!\!\! \equiv\  p(S_0,\delta_0) - 
\int_0^{S_0} p(S_0,\delta_0|S_1,\delta_1)\,f(S_1,\delta_1)\,{\rm d}S_1.
\end{eqnarray}
If $\delta_0\ge\delta_1$, then equation~(\ref{pdfs}) shows that 
$q=0$.  When $\delta_0<\delta_1$, then the integral above can 
be solved to yield equation~(\ref{qchandra}).  
Bond et al. (1991) discuss Chandrasekhar's derivation of 
$q(\delta_1,\delta_0,V_0)$.  Their discussion of excursion 
set trajectories associated with Gaussian random fields shows, 
with no calculation, that the expression above is correct.  

The excursion set approach of Bond et al. (1991) also shows why 
equation~(\ref{feq10}) must be correct.  Consider the set of all 
excursion set trajectories, and label each trajectory by its 
value of $\delta(V_0)\equiv\delta_0$ on scale $V_0$.  
Now, $q(\delta_1,\delta_0,V_0)$ gives the probability that such a 
trajectory lies below $\delta_1$ for all $V>V_0$, and 
$f(M_1,\delta_1|M_0,\delta_0)$ of equation~(\ref{fm1m2}) gives the 
fraction of trajectories that first cross the value $\delta_1$ on 
scale $V_1$, given that they have value $\delta_0$ on scale $V_0$.  
Integrating the product of these two expressions over all 
$\delta_0\le\delta_1$ gives the fraction of trajectories that first 
cross the value $\delta_1$ on the scale $V_1$, which is the same as 
equation~(\ref{fm}).  The extra factor of $M_0/M_1$ on the left 
hand side above is $\bar\rho V_0/M_1$ when written on the right 
hand side, which is consistent with equation~(\ref{nm1}).  

Expressions for the mean bias between haloes and mass can 
be obtained by taking similar limits.  
A little algebra shows that the peak background split of 
equation~(\ref{pbsplit}) could have been obtained directly from 
the corresponding Poisson limit, equation~(\ref{bign}).  

Expressions for the cross correlation between haloes and mass 
transform similarly, as well as for the higher order moments of 
the halo distribution all transform similarly.  For example, 
equation~(\ref{mualpha}) could have been derived by taking the 
limit of equation~(\ref{musl98}), etc.  

\section{Solution to the recursion relation}\label{solvrec}
This Appendix shows, by direct substitution, that 
equation~(\ref{fjk}) for $F(j|k)$ in the main text solves the 
recursion relation given in equation~(\ref{recur}).  

Equation~(\ref{recur}) can be rearranged to read 
\begin{eqnarray}
&&\!\!\!\!\!\!\!\!\!\!\!\!\!\!\!
\sum_{m>j}^k F(m,b_1|k,b_2)\,p(j,\,_1\!V_j|m,\,_1\!V_m) \nonumber \\ &=& 
{p(j,\,_1\!V_j,k,\,_2\!V_k)\,f^{\rm e}(\delta_2|k,V_k)\over F(k,b_2)} 
- F(j,b_1|k,b_2).
\label{rearr}
\end{eqnarray}
Equation~(\ref{fjb}) shows that the right hand side of this 
expression is 
\begin{equation}
{\rm RHS} = 
{p(j,\,_1\!V_j)\ p(k-j,\,_2\!V_k-\,_1\!V_j)\over p(k,\,_2\!V_k)} 
- F(j,b_1|k,b_2),
\end{equation}
where all the $p(n,V)$s are Poisson, so they are given by 
equation~(\ref{pois}).  If equation~(\ref{fjk}) for $F(j|k)$ is 
correct, then 
\begin{equation}
{\rm RHS} = 
(k-j)\,{k\choose j}\,{j^j\over k^k}\,\left({b_1\over b_2}\right)^{j+1}
\left(k-j{b_1\over b_2}\right)^{k-j-1}.
\label{rhs}
\end{equation}
Substituting equation~(\ref{pjbinom}) for $p(j,\,_1\!V_j|m,\,_1\!V_m)$ 
in the left hand side gives 
\begin{eqnarray}
&&\!\!\!\!\!\!\!\!\!\!\!\!\!
k\left(1-{b_1\over b_2}\right)\,{k\choose m}\,{m^m\over k^k}\,
\left(b_1\over b_2\right)^m\left(k-m{b_1\over b_2}\right)^{k-m-1}
\nonumber \\ &&
\times\ {m-1\choose j-1}\left({j\over m}\right)^{j-1}
\left(1 - {j\over m}\right)^{m-j} 
\end{eqnarray}
summed over all $j<m\le k$.  This reduces to 
\begin{eqnarray}
&&\!\!\!\!\!\!\!\!\!\!\!\!\!
(k-j)\,{k\choose j}\,{j^j\over k^k}\,\left({b_1\over b_2}\right)^{j+1}
k\left(1 - {b_1\over b_2}\right) \nonumber \\
&&\!\!\!\!\!\!\!\! \times 
\sum_{n=0}^N {N\choose n}\left({b_1\over b_2} + n{b_1\over b_2}\right)^n
\left(k - (j+1){b_1\over b_2} - n{b_1\over b_2}\right)^{N-n-1}
\end{eqnarray}
where $N=(k-j-1)$.  
Abel's generalization of the binomial theorem
\begin{equation}
(x+y)^N = \sum_{m=0}^N {N\choose m} x\,(x-mz)^{m-1}(y+mz)^{N-m},
\end{equation}
with $m = N-n$, $x = k(b_2-b_1)/b_2$, $y = (k-j)(b_1/b_2)$, 
and $z = -(b_1/b_2)$, reduces this to equation~(\ref{rhs}).  

A similar recursion relation is satisfied by $f(j,b_1|k,b_2)$.  
Namely, 
\begin{eqnarray}
&&\!\!\!\!\!\!\!\!\!\!\!\!\!\!\!
\sum_{m>j}^k f(m,b_1|k,b_2)\,p(j,\,_1\!V_j|m,\,_1\!V_m) \nonumber \\ &=& 
{p(j,\,_1\!V_j,k,\,_2\!V_k)\,f^{\rm e}(\delta_2|k,V_k)\over f(k,b_2)} 
- f(j,b_1|k,b_2).
\label{recurf}
\end{eqnarray}
Since now trajectories are known to be centred on particles, 
\begin{eqnarray}
p(j,\,_1\!V_j,k,\,_2\!V_k) &=& 
p(j-1,\,_1\!V_j)\ p(k-j,\,_2\!V_k-\,_1\!V_j) \nonumber \\
&=& {p(j,\,_1\!V_j)\over b_1}\ p(k-j,\,_2\!V_k-\,_1\!V_j).  
\end{eqnarray}
Since $F(k,b_2) = b_2\,f(k,b_2)$ (equation~\ref{effs}), the right 
hand side of equation~(\ref{recurf}) is $(b_2/b_1)$ 
times that in equation~(\ref{rearr}).  

Similarly, since now trajectories are centred on particles,  
\begin{equation}
p(j,\,_1\!V_j|m,\,_1\!V_m) = 
{p(j-1,\,_1\!V_j)\ p(m-j,\,_1\!V_m-\,_1\!V_j)\over p(m-1,\,_1\!V_m)}.
\end{equation}
This is the same as equation~(\ref{pjbinom}).  Therefore, if the 
left hand side of equation~(\ref{recurf}) is to equal $(b_2/b_1)$ 
times the left hand side of equation~(\ref{rearr}), then it must be 
that $f(j,b_1|k,b_2) = (b_2/b_1)\,F(j,b_1|k,b_2)$.  This is just what 
is required by equation~(\ref{efjs}).  Thus, if $f(j|k)$ is given 
by equation~(\ref{efjs}), then it satisfies the recursion 
relation~(\ref{recurf}).  

\section{Averaging over all volumes}\label{bower}
This Appendix shows that the expressions for the conditional 
and unconditional mass functions are obtained by an averaging 
process envisaged by Bower (1991).  Namely, the averaging 
is over all possible subvolumes, not necessarily connected, 
that are contained entirely within a parent volume.  

Suppose space is divided up into a large number $C$ of 
infinitesimally small cells, each of volume $v$.  
The cells are sufficiently small that each cell is either empty, 
or it contains one and only one particle.  Suppose that there 
are $N$ particles distributed in this space.  This means that 
$N$ of the $C$ cells are occupied.  Now choose $c$ cells in random 
order without replacement from the total set of $C$ cells.  
The probability that $n$ of these $c$ cells are occupied is 
\begin{eqnarray}
p(n,c) &=& {c\choose n}\ \times 
{N(N-1)\cdots(N-n+1)\over C(C-1)\cdots (C-n+1)}\nonumber \\ 
&&\qquad\times {(C-N)\cdots(C-N-(c-n)+1)\over (C-n)\cdots(C-c+1)}\nonumber\\
&=& {{N\choose n}{C-N\choose c-n}\over {C\choose c}}.
\label{nchoozc}
\end{eqnarray}
When $C\gg c\gg N\gg n$, Stirling's approximation for all the 
factorials except $n!$ reduces this to 
\begin{displaymath}
p(n,c) \to {1\over n!}\left({cN\over C}\right)^n {\rm e}^{-cN/C} .
\end{displaymath}
Now, $Cv$ is the total volume, so $(N/Cv)$ is the average 
number density of particles; denote it by $\bar n$.  The 
parameter $cv$ is the size of the cell made of $c$ infinitesimal 
cells; set $cv\equiv V$.  Then $(cN/C) = \bar nV$ and this 
expression is the same as equation~(\ref{pois}).  
This shows explicitly how the Poisson distribution is obtained 
by choosing, in random order without replacement, a series of 
volume elements of the total space, and weighting each series 
of choices with the probability that it occurs.  
Since $F(j,b)$ is simply the product of $p(j,V_j)$ with 
a quantity that depends on $b$ but not $V$, the argument 
above applies to $F(j,b)$ also.  In particular, since the 
volume elements $c$ are chosen at random from the full space, 
there is no requirement that they be adjacent.  

A similar argument can be used to derive equation~(\ref{pjoint}).  
Namely, suppose $V_2$, containing exactly $N$ particles is divided 
up into a large number $C$ of small volumes $v$.  Then, the 
probability that in $c$ volumes, chosen randomly without 
replacement from $C$, there are exactly $n$ occupied volumes, 
when it is known that there are exactly $N$ occupied volumes in 
$C$, is given by the same expression (\ref{nchoozc}) as before.  
When $C\gg c\gg N\ge n$, Stirling's approximation for all the 
factorials except the ${N\choose n}$ term reduces this to 
\begin{equation}
p(n,c|N,C) \to {N\choose n}\ \left({c\over C}\right)\ 
\left(1 - {c\over C}\right)^{N-n} .
\end{equation}
With $cv\equiv V_1$, this is the same as equation~(\ref{pjoint}), 
since $Cv\equiv V_2$.  Again, the only constraint on the volume 
elements $c$ is that they lie entirely within $V_2$.  There is 
no requirement that they be adjacent.

What remains to be shown is that $F(j|k)$ is also obtained by 
a sampling process in which the different volume elements which 
make up $_1\!V_j$ are chosen randomly without replacement from 
$_2\!V_k$, so, in particular, they are not necessarily adjacent 
to each other.  This follows from the original derivation, or 
from the fact that the derivative of $p(j|k)$ is so easily related 
to $F(j|k)$, or from the derivation of $f(j|k)$ given in 
Sheth (1995).  

\end{document}